\documentclass[aps,prb,twocolumn,superscriptaddress,showpacs,showkeys,amsmath,amssymb]{revtex4}

\usepackage{epsfig,amsmath,amssymb,color}
\usepackage[T1]{fontenc}
\usepackage[latin9]{inputenc}
\usepackage{epstopdf}
\renewcommand{\theequation}{\arabic{section}.\arabic{equation}}

\begin{document}

\title{Hubbard models with nearly flat bands:\\ 
       Ground-state ferromagnetism driven by kinetic energy}

\author{Patrick M\"{u}ller}
\affiliation{Institut f\"{u}r theoretische Physik,
          Otto-von-Guericke-Universit\"{a}t Magdeburg,
          P.O. Box 4120, 39016 Magdeburg, Germany}

\author{Johannes Richter}
\affiliation{Institut f\"{u}r theoretische Physik,
          Otto-von-Guericke-Universit\"{a}t Magdeburg,
          P.O. Box 4120, 39016 Magdeburg, Germany}

\author{Oleg Derzhko}
\affiliation{Institute for Condensed Matter Physics,
          National Academy of Sciences of Ukraine,
          Svientsitskii Street 1, 79011 L'viv, Ukraine}
\affiliation{Department for Theoretical Physics,
          Ivan Franko National University of L'viv,
          Drahomanov Street 12, 79005 L'viv, Ukraine}
\affiliation{Institut f\"{u}r theoretische Physik,
          Otto-von-Guericke-Universit\"{a}t Magdeburg,
          P.O. Box 4120, 39016 Magdeburg, Germany}
\affiliation{Abdus Salam International Centre for Theoretical Physics,
          Strada Costiera 11, 34151 Trieste, Italy}

\date{\today}

\begin{abstract}
We consider the standard repulsive Hubbard model with  a
flat lowest-energy band for two one-dimensional lattices
(diamond chain and ladder)
as well as for a two-dimensional lattice
(bilayer) at half filling of the flat band. 
The considered models do not fall in the class of Mielke-Tasaki flat-band
ferromagnets, since they do not obey the connectivity conditions.
However, the ground-state ferromagnetism can emerge,
if the flat band becomes dispersive.
To study this kinetic-energy-driven ferromagnetism 
we use perturbation theory and exact diagonalization of finite lattices.
We find as a typical scenario 
that small and moderate dispersion may lead to a ferromagnetic ground state for sufficiently large on-site Hubbard repulsion $U>U_c$, 
where $U_c$ increases monotonically with the acquired bandwidth. 
However, we also observe for some specific parameter cases, 
that (i) ferromagnetism appears at already very  small $U_c$, 
(ii) ferromagnetism does not show up at all, 
(iii) the critical on-site repulsion $U_c$ is a nonmonotonic function of the bandwidth, 
or that (iv) a critical bandwidth is needed to open the window for ground-state ferromagnetism.
\end{abstract}

%    71.27.+a Strongly correlated electron systems; heavy fermions
%    71.10.Fd Lattice fermion models (Hubbard model, etc.)


\maketitle

\section{Introductory remarks}
\label{sec1}
\setcounter{equation}{0}

Explaining ferromagnetism from a simple model of itinerant electrons such as the standard Hubbard model 
is a long-standing problem in the condensed matter theory.
Among many routes leading to ferromagnetism 
the so-called flat-band ferromagnetism of Mielke and Tasaki\cite{mielke,tasaki,rev} is of special interest.
On the one hand, 
many results for Mielke-Tasaki flat-band ferromagnetism have been obtained rigorously.
On the other hand, 
this mechanism is important for material design, 
since it opens interesting possibilities to obtain ferromagnetic materials in which magnetic atoms are completely missing.
In brief, the mechanism of this kind of ferromagnetism looks as follows.\cite{mielke,tasaki,rev}
Flat-band ground states 
(i.e., the one-particle states from completely dispersionless band which is the lowest-energy one)
can be considered as one-particle states which are localized within small trapping cells on a lattice.\cite{lm,le,others_le}
Therefore, 
exact many-electron ground states at low electron densities can be constructed simply by filling the traps.
Importantly, 
in the case of connected (overlapping) traps,
electrons being in symmetric spin states avoid the on-site Hubbard
repulsion,  
and, as a result, these states remain within the ground-state manifold for
$U>0$ with a $U$-independent energy.
Thus, the (degenerate) ground state consists of a set of ferromagnetic clusters.
If the electron density exceeds a threshold value, 
a macroscopic wrapping ferromagnetic cluster appears
and ferromagnetism dominates the ground-state properties of thermodynamically large systems.\cite{mielke,tasaki,rev,le,myk1,myk2,ijmpb}
This ferromagnetism is robust against perturbation,
i.e., the ferromagnetic state remains stable for slightly perturbed models 
which have a moderate change in the hopping integrals 
leading to a slightly dispersive one-electron band.\cite{tasaki_jsp,kusakabe}

The above description of the emergence of ground-state ferromagnetism is based on the assumption, 
that the trapping cells have common sites,
i.e., the so-called connectivity condition is satisfied for the localized one-electron states.
In other words,
the localized states overlap
and this was essential for the proofs in Refs.~\onlinecite{mielke,tasaki}.
On the other hand,
there are lattices which have lowest-energy flat bands but the traps do not have common sites
(nonoverlapping or isolated traps).
Those flat-band lattices
cannot support the above described mechanism for ferromagnetism, 
since the trapped electrons cannot be in contact with each other, and, thus are unable to correlate.
Hence,  flat-band Hubbard models with isolated traps do not exhibit ferromagnetism at zero temperature, 
rather  there is a macroscopically degenerate (i.e., the degeneracy
grows exponentially with the system size) ground-state manifold, where
paramagnetic states dominate.\cite{batista,prb2009,epjb2011,mielke2012}
However, the macroscopically degenerate ground-state
manifold is very sensitive to small perturbations which may lead  to subtle
effects of violations of the flat-band conditions.
This scenario has been investigated in Ref.~\onlinecite{oleg-johannes} 
for the specific example of the frustrated diamond chain.
It was demonstrated
that the macroscopically degenerate ground-state manifold with all traps filled by electrons
results in a non-magnetic zero-temperature phase,\cite{oleg-johannes,prb2009}
but small deviations from the ideal flat-band geometry of hopping integrals 
(which makes the flat band slightly dispersive)
lead to a fully polarized ferromagnetic many-electron ground state if $U>U_c$.
The value of $U_c$ depends on the strength of the deviation from the ideal geometry. 
Note that another route to ground-state ferromagnetism without 
connectivity condition in the flat band was discussed in
Ref.~\onlinecite{gkg}.

In the present paper we broaden and generalize our previous study 
on the dispersion-driven ferromagnetism in flat-band Hubbard systems.\cite{oleg-johannes}
As already mentioned above, those studies referred to one particular lattice, 
namely to an azurite-like\cite{azurite} diamond-Hubbard chain. 
Moreover, analytical calculations presented in Ref.~\onlinecite{oleg-johannes} 
were restricted to the fourth-order perturbation theory for a two-cell chain.
In the present study
we extend the analytical calculations to higher-orders perturbation theory
this way validating the previous results.
More importantly,
we consider other lattices with isolated trapping cells, 
the one-dimensional ladder and the two-dimensional bilayer. These new
lattices have more degrees of freedom to constitute 
 deviations from the ideal flat-band geometry.
Thus, we will demonstrate that the dispersion-driven ferromagnetism is a rather general mechanism 
to establish ferromagnetic ground states in Hubbard models 
having isolated trapping cells in the flat-band limit.
In addition to the analytical perturbation theory, 
we also perform extensive exact-diagonalization studies.
Our analysis will, on the one hand, confirm the conclusions derived from the
study of the Hubbard diamond chain.\cite{oleg-johannes}
On the other hand, we will discuss further consequences 
of deviations from the ideal flat-band geometry on ferromagnetism.
In particular, 
we find that in some cases the required threshold on-site repulsion $U_c$ may be quite
small, 
whereas in other cases ferromagnetic ground states do not appear at all.
There are also cases when ferromagnetic ground states appear only, 
if the acquired bandwidth exceeds a threshold, 
and then $U_c$ becomes a nonmonotonic function of the bandwidth.
Our findings are compactly collected in phase diagrams, 
obtained both by analytical treatment and exact diagonalization, 
which indicate the regions of dispersion-driven ground-state ferromagnetism.

The paper is organized as follows.
After a brief description of the models to be considered (Sec.~\ref{sec2}) and the methods to be used (Sec.~\ref{sec3})
we pass to a discussion of the obtained results for the diamond chain (Sec.~\ref{sec4}), the ladder (Sec.~\ref{sec5}), 
and the bilayer (Sec.~\ref{sec6}). 
We briefly summarize our results in Sec.~\ref{sec7}.
Several appendices present some lengthy formulas which are relevant for
the discussion in the main text of the paper.

\section{Models}
\label{sec2}
\setcounter{equation}{0}

\begin{figure}%[h]
\begin{center}
\includegraphics[clip=on,width=72mm,angle=0]{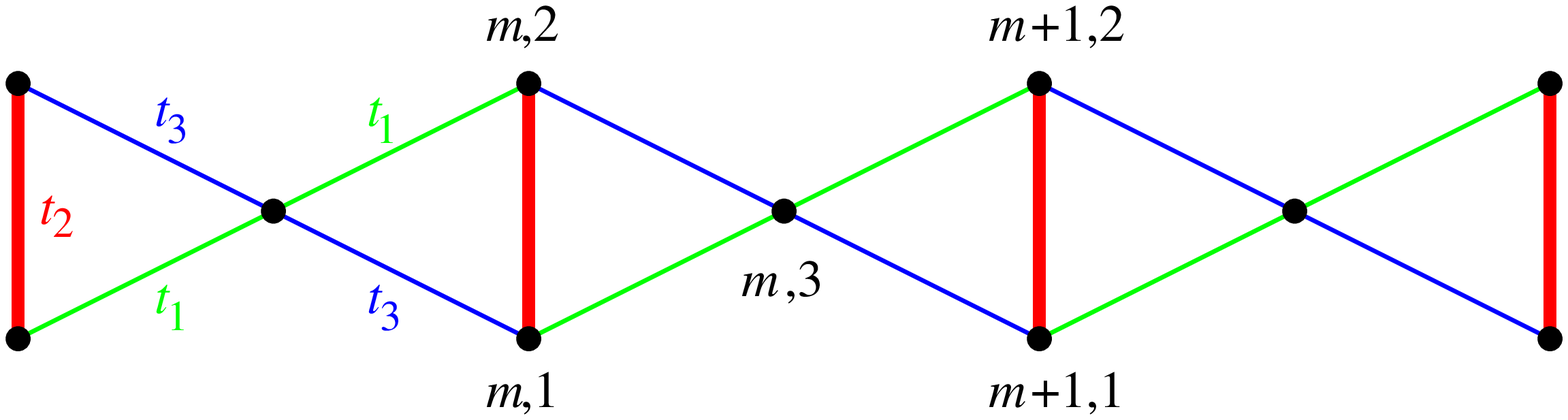}\\
\vspace{5mm}
\includegraphics[clip=on,width=65mm,angle=0]{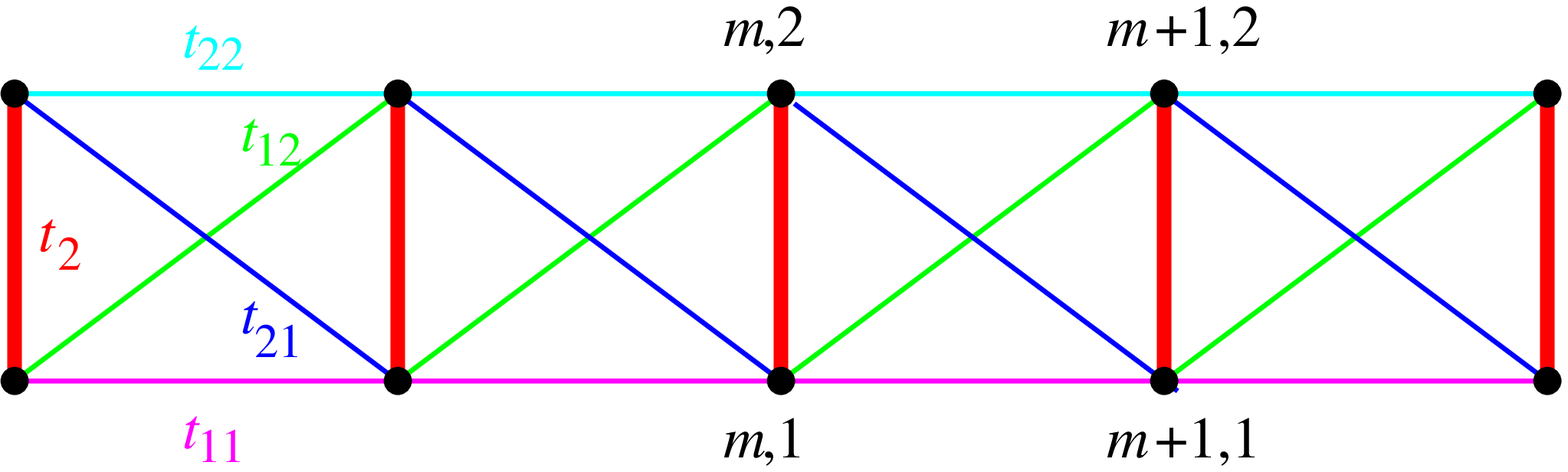}\\
\vspace{5mm}
\includegraphics[clip=on,width=75mm,angle=0]{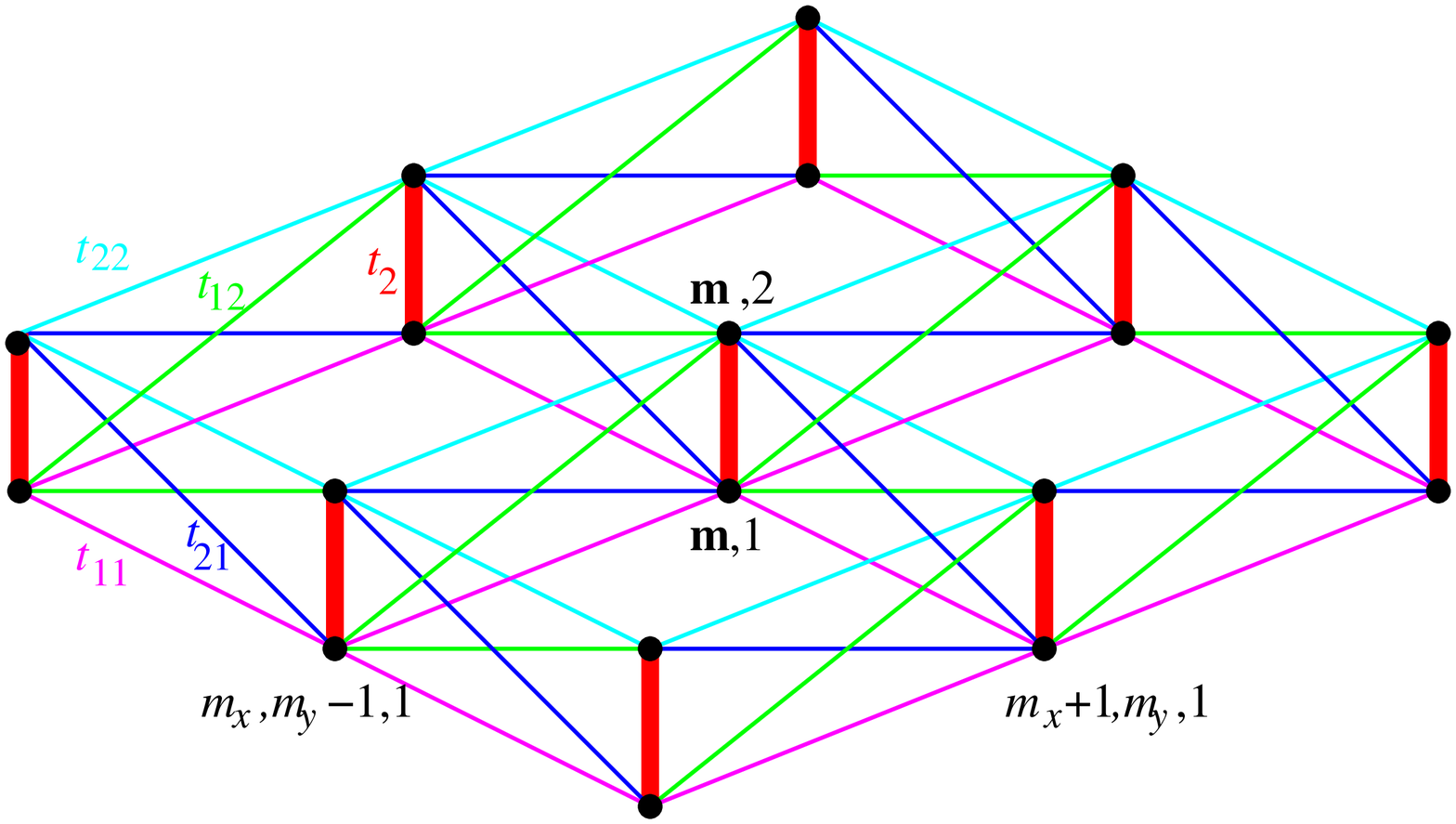}\\
\vspace{5mm}
\caption{(Color online)
Lattices considered in the present paper:
The frustrated diamond chain, the frustrated two-leg ladder, and the frustrated bilayer
(from top to bottom).
The sites are enumerated by two indexes $m,i$:
The first one enumerates the cells, 
$m=1,\ldots,{\cal{N}}$,
and 
the second one enumerates the sites within a cell,
$i=1,2,3$ (diamond) and $i=1,2$ (ladder and bilayer).
The hopping integral for the vertical bond is $t_2$,
whereas the hopping integral along the bond connecting the sites $m,i$ and $m+1,j$ is denoted by $t_{ij}$,
see also the main text.
For ideal flat-band geometry $t_{ij}=t$ and $2t<t_2$ (diamond and ladder) or $4t<t_2$ (bilayer).}
\label{f01}
\end{center}
\end{figure}

We consider the standard repulsive one-orbital Hubbard model with the Hamiltonian
\begin{eqnarray}
\label{201}
H=\sum_{\sigma=\uparrow,\downarrow}H_{0,\sigma}+H_U,
\nonumber\\
H_{0,\sigma}=\sum_{(ij)}t_{ij}\left(c_{i,\sigma}^\dagger c_{j,\sigma}+c_{j,\sigma}^\dagger c_{i,\sigma}\right),
\;\;\;
t_{ij}>0,
\nonumber\\
H_U=U\sum_{i}n_{i,\uparrow}n_{i,\downarrow},
\;\;\;
U>0,
\end{eqnarray}
where generally accepted notations are used in Eq.~(\ref{201}).
We investigate the Hubbard model (\ref{201}) on two one-dimensional and one two-dimensional $N$-site lattices 
which are shown in Fig.~\ref{f01},
namely the frustrated diamond chain, the frustrated two-leg ladder, and the frustrated bilayer.
In case of ideal flat-band geometry all hopping integrals $t_{ij}=t$ are
equal, except the hopping integral on the vertical bond $t_2$. 
Then 
one of the one-electron bands is strictly flat and it becomes the lowest one, 
if $t_2$ is sufficiently large.
The localized-electron states are then located (trapped) on the vertical $t_2$-bonds.
Obviously, the trapping cells do not have common sites, the connectivity condition is violated, 
and the zero-temperature state in the subspaces with $n\le{\cal{N}}$ electrons are nonmagnetic. 
From Fig.~\ref{f01} it is obvious, 
that the number of trapping cells ${\cal{N}}$ for the diamond chain and the ladder/bilayer 
is ${\cal{N}}=N/3$ and ${\cal{N}}=N/2$, respectively. 

We consider  deviations from the ideal flat-band geometry of the following
form: For the diamond chain, following Ref.~\onlinecite{oleg-johannes}, we
set $t_{13}=t_{32}=t_1\ne t_{23}=t_{31}=t_3$, $t_1+t_3=2t<t_2$ (azurite-like
geometry;\cite{azurite} for more general deformations see
Ref.~\onlinecite{fnt}).  It is convenient to parameterize the azurite-like
distortion as follows: 
\begin{eqnarray} 
\label{202} 
t_1=t(1+\delta), 
\;\;\;
t_3=t(1-\delta); 
\nonumber\\ 
t=\frac{t_1+t_3}{2}, 
\;\;\;
\delta=\frac{t_1-t_3}{t_1+t_3}.  
\end{eqnarray} 
For the ladder/bilayer
$t_{11}$, $t_{12}$, $t_{21}$, and $t_{22}$ may be different, but we assume
$t_{11}+t_{12}+t_{21}+t_{22}=4t$ and $2t< t_2$ (ladder) or $4t< t_2$
(bilayer).  Again it is convenient to introduce the following
parameterization: 
\begin{eqnarray}
\label{203} 
t_{11}=t_l(1+\delta_l),
\;\;\; 
t_{12}=t_f(1+\delta_f), 
\nonumber\\ 
t_{21}=t_f(1-\delta_f), 
\;\;\;
t_{22}=t_l(1-\delta_l); 
\nonumber\\ 
t_l=\frac{t_{11}+t_{22}}{2}, 
\;\;\;
\delta_l=\frac{t_{11}-t_{22}}{t_{11}+t_{22}}, 
\nonumber\\
t_f=\frac{t_{12}+t_{21}}{2}, 
\;\;\;
\delta_f=\frac{t_{12}-t_{21}}{t_{12}+t_{21}} 
\end{eqnarray} 
with
$t_l+t_f=2t$.

In the distorted systems the lowest flat band with energy
$\varepsilon_1$ acquires a dispersion, i.e.,
$\varepsilon_1\to\varepsilon_1(\kappa)$,
resulting in a nonzero bandwidth $W_1>0$.
In Ref.~\onlinecite{oleg-johannes},
the acquired dispersion was characterized by a parameter $W_1/w_{2}$,
where $w_2$ denotes the bandwidth of the dispersive bands for the ideal flat-band geometry 
(note that for the diamond chain there are two dispersive bands with identical bandwidth).
Furthermore,
for the diamond chain we have
$W_1\approx 2(t_3-t_1)^2/t_2$,
$w_{2}\approx 2(t_3+t_1)^2/t_2$
and therefore
$W_1/w_{2}\approx\Omega^2$,
where  
$\Omega\equiv\left\vert(t_3-t_1)/(t_3+t_1)\right\vert$ used in Ref.~\onlinecite{oleg-johannes}
equals to $\vert\delta\vert$,
cf. Eq.~(\ref{202}).
However, 
since for the Hubbard ladder/bilayer 
the acquired bandwidth is not the only relevant parameter that controls the emergence of
ferromagnetism,
we prefer to use throughout this paper the above introduced parameters
$t$ and $\delta$ for the diamond chain
and
$t_l$, $t_f$, $\delta_l$, and $\delta_f$ for the ladder/bilayer.

\section{Methods}
\label{sec3}
\setcounter{equation}{0}

In our study we use an analytical perturbation-theory approach and numerical exact diagonalization.
Let us briefly explain these methods.
The starting point of the perturbation theory is  
the splitting of the Hamiltonian $H$ of the problem at hand into the main part (unperturbed Hamiltonian) ${\sf{H}}_0$ 
and the perturbation ${\sf{V}}$,
i.e., $H={\sf{H}}_0+{\sf{V}}$.
Then we use the perturbation-theory formulas given in Ref.~\onlinecite{klein} (see also Appendix~A) to
determine the influence of the perturbation  ${\sf{V}}$ on the degenerate ground-state
manifold.
Since $t_2>0$ is the largest hopping integral and $U>0$,
the main part consists of the hopping terms on the vertical bonds and all on-site repulsion terms.
The perturbation consists of all other hopping terms.
Next we have to find all eigenstates and eigenvalues of the unperturbed Hamiltonian ${\sf{H}}_0$.
For $N$ sites and $n$ electrons there are altogether ${\cal{C}}_{2N}^n=(2N)!/[n!(2N-n)!]$
eigenstates.
For example, for $n={\cal{N}}=2,\,3,\,4,\,5$ ladder problems we have 28, 220, 1820, 15504 eigenstates, respectively.
In the considered regime, i.e., dominating positive $t_2$, $U>0$ is sufficiently large, and
$n={\cal{N}}$,
the ground state is $2^n$-fold degenerate,
i.e., 4-, 8-, 16-, 32-fold degenerate for $n={\cal{N}}=2,\,3,\,4,\,5$.
It has the form:
\begin{eqnarray}
\label{301}
\vert{\rm{GS}}\rangle=l^\dagger_{1,\sigma_1}\ldots l^\dagger_{n,\sigma_n}\vert{\rm{vac}}\rangle,
\nonumber\\
l^\dagger_{m,\sigma_m}
=\frac{1}{\sqrt{2}}\left(c^\dagger_{m,1,\sigma_m}-c^\dagger_{m,2,\sigma_m}\right).
\end{eqnarray}
The choice of the concrete linear combinations of states (\ref{301})
used as a starting point of perturbation theory
is related to the model with
perturbation.
Supposing an effective magnetic Heisenberg model for the low-energy degrees
of freedom,\cite{oleg-johannes} the choice   
of ground states of the unperturbed Hamiltonian ${\sf{H}}_0$ which account
the SU(2) symmetry 
of the Hubbard Hamiltonian is straightforward, for more details see
Appendix~B.
The resulting perturbation-theory formulas up to the sixth order are collected in Appendix~A
(see also Appendices~C, D, and E).
It is in order to mention here,
that in the small-$U$ limit, in addition to the states (\ref{301}), also states with two electrons in one cell,
become relevant. 
As a result, the perturbation theory starting from the set of states (\ref{301}) may fail for $U \to 0$, see below.

To perform the fourth and sixth order perturbation theory 
we use the symbolic computation software {\it Mathematica}. 
To implement the symbolic calculation we used the SNEG package, 
see Ref.~\onlinecite{sneg},
for Mathematica. 
The package handles the non-commutative multiplication of, 
e.g., 
fermionic creation and annihilation operators. 
This is required to perform the perturbation theory in higher order for larger Hubbard
clusters. 
For a compact sketch of the procedure see Appendix~F.

For the numerical exact diagonalization we use  J.~Schulenburg's {\it spinpack}.\cite{spinpack1,spinpack2}
This code allows the calculation of the ground state for the Hubbard model with a half-filled lowest band
up to $N=20$ sites.
Thus, by considering various system sizes the finite-effects can be estimated. 
The comparison of the results obtained by  two different approaches 
finally allows to get a consistent description of the ground-state phases 
of the considered Hubbard systems.

\section{Diamond chain}
\label{sec4}
\setcounter{equation}{0}

The Hubbard model Hamiltonian on the diamond chain is given in Eq.~(\ref{201}) 
with the following explicit form for $H_{0,\sigma}$:
\begin{eqnarray}
\label{401}
H_{0,\sigma}=\sum_{m}\left[
t_{2}c_{m,1,\sigma}^\dagger c_{m,2,\sigma}
\right.
\nonumber\\
\left.
+t_{1}\left(c_{m,1,\sigma}^\dagger c_{m,3,\sigma}+c_{m,3,\sigma}^\dagger c_{m+1,2,\sigma}\right)
\right.
\nonumber\\
\left.
+t_{3}\left(c_{m,2,\sigma}^\dagger c_{m,3,\sigma}+c_{m,3,\sigma}^\dagger c_{m+1,1,\sigma}\right)
+{\rm{H.c.}}\right],
\end{eqnarray}
see Fig.~\ref{f01}.
Eq.~(\ref{401}) corresponds to an azurite-like deformation.\cite{azurite}
Furthermore, we assume half filling of the lowest nearly flat one-electron band,
i.e., the number of electrons equals the number of cells $n={\cal{N}}$.

Extensive exact-diagonalization calculations for this model were reported in
Ref.~\onlinecite{oleg-johannes}.
However,
the analytical treatment by perturbation theory was restricted to fourth-order
calculations 
for the two-cell diamond chain with open boundary conditions consisting of
$N=5$ sites. (Note, that for the special diamond-chain geometry
the second-order perturbation theory is not sufficient to describe
ground-state ferromagnetism.\cite{oleg-johannes})  
 In this paper we present  the sixth-order perturbation theory  and consider also a
larger cluster  consisting of three cells in fourth-order perturbation theory.
That allows to validate the previous lower-order approach and promises a
better agreement with  exact diagonalization for larger deviations from the
ideal flat-band geometry.

\begin{figure}%[h]
\begin{center}
\includegraphics[clip=on,width=62.5mm,angle=0]{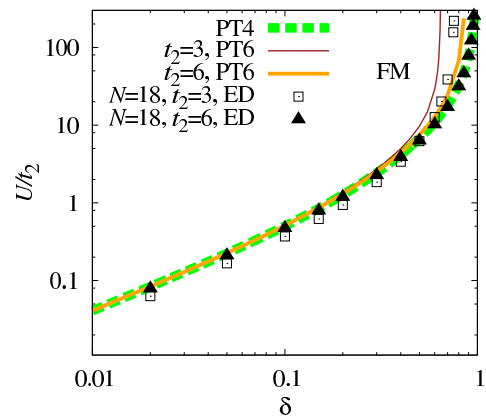}
\caption{(Color online)
Phase diagram for the Hubbard diamond chain.
Ferromagnetic ground states appear for $U>U_c$.
$U_c$ is shown as a function of $\delta$, $t=1$, see Eq.~(\ref{202}).
The various critical lines $U_c(\delta)$ are obtained by perturbation theory and exact diagonalization.}
\label{f02}
\end{center}
\end{figure}

The results 
for the triplet and singlet energies calculated for the cluster of $N=5$ sites with $n=2$ electrons up to the sixth order,
\begin{eqnarray}
\label{402}
E_t=-2t_2+E^{(2)}+E_t^{(4)}+E_t^{(6)}+\ldots,
\nonumber\\
E_s(U)=-2t_2+E^{(2)}+E_s^{(4)}(U)+E_s^{(6)}(U)+\ldots,
\end{eqnarray}
are given in Appendix~C.
From the obtained data one can see 
that with increasing of the order of perturbation-theory calculations
the analytical results for the triplet and singlet energies monotonically approach the exact-diagonalization data from above.
The critical on-site repulsion 
 $U_c$ is determined from the equation $E_t=E_s(U_c)$.
In fourth-order perturbation-theory we get a compact formula\cite{oleg-johannes}
\begin{eqnarray}
\label{403}
\frac{U^{(4)}_c}{t_2}=\frac{\sqrt{16+65\delta^{2}}+9\vert\delta\vert}{1-\delta^{2}}\vert\delta\vert.
\end{eqnarray}
Eq.~(\ref{403}) implies that in fourth order $U_c/t_2$ depends only on the deviation from
the ideal flat-band geometry controlled by $\delta$,
but not on $t$ or $t_2$.
Unfortunately,
in sixth order $U^{(6)}_c$ obtained as a solution of the equation
$E_t^{(4)}+E_t^{(6)}=E_s^{(4)}(U^{(6)}_c)+E_s^{(6)}(U^{(6)}_c)$
has to be calculated numerically, and cannot be presented in a compact
analytical form.
By contrast to  $U^{(4)}_c$, the sixth-order result  $U^{(6)}_c/t_2$ weakly
depends on $t_2$, which was also found in our exact-diagonalization results.
The corresponding results for $U^{(4)}_c$ and $U^{(6)}_c$ are shown in
Fig.~\ref{f02}. 
It is evident, that the difference between the values of $U_c^{(4)}$ and $U_c^{(6)}$ at least for small $\delta$, where the perturbation theory is valid, is small
(the difference in Fig.~\ref{f02} becomes only visible if $\delta$ exceeds 0.4).
Thus, we confirm that the simple equation (\ref{403}) describes the phase
boundary surprisingly well.

Another way to extend the previous perturbation-theory calculations of
Ref.~\onlinecite{oleg-johannes} is to enlarge the cluster sizes used for the
perturbation theory. For that we consider $n=3$ electrons on the three-cell diamond chain with open boundary conditions which has $N=8$ sites.
Already in fourth order the perturbation theory becomes more ambitious,
since we have to take into account much more states, see Appendix~C.
Remarkably, for the larger cluster we get the same value of $U^{(4)}_c$ as given in Eq.~(\ref{403}).

Our results  are summarized in Fig.~\ref{f02},
where we also show some exact-diagonalization results obtained earlier.\cite{oleg-johannes}
This figure provides evidence, that
the sixth-order perturbation-theory calculations ($N=5$) almost do not change the predictions for $U_c(\delta)$ according to Eq.~(\ref{403}),
although there is a weak dependence of $U_c/t_2$ on $t_2$ in agreement with exact-diagonalization data
(compare the curves PT6 for $t_2=3$ and $t_2=6$ in Fig.~\ref{f02}).
The fact that Eq.~(\ref{403}) has been obtained now from calculations for both two-cell and three-cell diamond chains
(i.e., for $N=5$ and $N=8$), also  
explains the  good agreement of Eq.~(\ref{403}) with exact-diagonalization results for longer chains 
(e.g., for ${\cal{N}}=6$ cells, see Fig.~\ref{f02}).
Finally,
we emphasize again that our new results demonstrate that the formula for
$U_c$ given in Eq.~(\ref{403}) provides a simple and sufficiently precise criteria 
for emergence of ground-state ferromagnetism in the Hubbard diamond chain.

\section{Ladder}
\label{sec5}
\setcounter{equation}{0}

Next we consider as a new example for a flat-band model with isolated
trapping cells the Hubbard model on a frustrated ladder, see
Fig.~\ref{f01}.
We point out at the beginning that, by contrast to the diamond chain, there is
no intermediate site between two trapping cells.
The explicit form for $H_{0,\sigma}$ in Eq.~(\ref{201}) is
\begin{eqnarray}
\label{501}
H_{0,\sigma}=\sum_{m}\left(
t_{2}c_{m,1,\sigma}^\dagger c_{m,2,\sigma} \qquad \qquad
\right.
\nonumber\\
\left.
+t_{11}c_{m,1,\sigma}^\dagger c_{m+1,1,\sigma}
+t_{12}c_{m,1,\sigma}^\dagger c_{m+1,2,\sigma}
\right.
\nonumber\\
\left.
+t_{21}c_{m,2,\sigma}^\dagger c_{m+1,1,\sigma}
+t_{22}c_{m,2,\sigma}^\dagger c_{m+1,2,\sigma}
+{\rm{H.c.}}\right), 
\end{eqnarray}
see Fig.~\ref{f01}.

Using the notations of Eq.~(\ref{203}),
the one-electron dispersion relations for this model can be written in a compact manner as follows:
\begin{eqnarray}
\label{502}
\varepsilon_{1,2}(\kappa)
=2t_l\cos\kappa \qquad \qquad \qquad \qquad \qquad \qquad 
\nonumber\\
\mp\sqrt{\left(t_2+2t_f\cos\kappa\right)^2+ 4t_l^2\delta_l^2\cos^2\kappa+4t_f^2\delta_f^2\sin^2\kappa}.
\end{eqnarray}
Flat-band geometry occurs 
when $t_{11}=t_{12}=t_{21}=t_{22}=t$ or $t_l=t_f=t$, $\delta_l=\delta_f=0$ and $2t<t_2$.
Then 
$\varepsilon_{1}(\kappa)=\varepsilon_{1}=-t_2$
and
$\varepsilon_{2}(\kappa)=t_2+4t\cos\kappa > \varepsilon_{1}$.

We consider a quite general deviation from the ideal flat-band geometry, 
and assume only that $t_{11}+t_{12}+t_{21}+t_{22}=4t$ or $t_l+t_f=2t$ and $2t<t_2$.
Thus after fixing $t_l$ and $t_f$ with the restriction $t_l+t_f=2t<t_2$
we are left with two free parameters, $\delta_l$ and $\delta_f$ [see
Eq.~(\ref{203})], constituting a two-dimensional  parameter region.
Except the general case of deformations, we will also
consider two special deformations,
(i) a symmetric deformation 
with $t_{11}=t_{22}$, $t_{12}=t_{21}$ and $t_{11} \ne t_{12}$ ($t_l\ne t_f$, $\delta_l=\delta_f=0$)
and
(ii) 
a semi-symmetric deformation 
with $t_{11}=t_{12}$, $t_{21}=t_{22}$ and $t_{11} \ne t_{21}$ ($t_l=t_f=t$, $\delta_l=\delta_f=\delta\ne
0$) which is identical to  
$t_{11}=t_{21}$, $t_{12}=t_{22}$ and $t_{11} \ne t_{12}$ ($t_l=t_f=t$, $\delta_l=-\delta_f=\delta\ne 0$),  
since all results depend only on $\delta_{l}^2$ and $\delta_{f}^2$, 
see, e.g., Eq.~(\ref{502}). 
For case (i) the dispersion relation  Eq.~(\ref{502}) becomes
\begin{eqnarray}
\label{503}
\varepsilon_{1,2}(\kappa)=\mp t_2 + 2\left(t_{l} \mp t_{f}\right)\cos\kappa,
\end{eqnarray}
whereas for case (ii) translates  into
\begin{eqnarray}
\label{504}
\varepsilon_{1,2}(\kappa) = 2t\cos\kappa\mp\sqrt{\left(t_2+2t\cos\kappa\right)^2+ 4t^2\delta^2}.
\end{eqnarray}
It is worth noting
that the acquired bandwidth of the former flat band due to the symmetric deformation may be larger than due to the semi-symmetric one.
On the other hand, 
while the symmetric deformation does not lead to ferromagnetic ground states
at all,
see below,
the semi-symmetric one produces ferromagnetic ground states for very small $U>U_c$,
see below.
Obviously, the acquired bandwidth as the only relevant
parameter is insufficient to characterize the capability to obtain ground-state ferromagnetism.

In what follows 
we first discuss perturbation-theory results in comparison with exact-diagonalization data for ladders up to ${\cal{N}}=4$ cells ($N=8$ sites)
and then present all analytical findings along with exact diagonalization for $N=12,\,16,\,20$ (${\cal{N}}=6,\,8,\,10$) in phase diagrams.

\subsection{Two electrons and two cells}
\label{ladder_two_cells}

\begin{figure}%[h]
\begin{center}
\includegraphics[clip=on,width=78mm,angle=0]{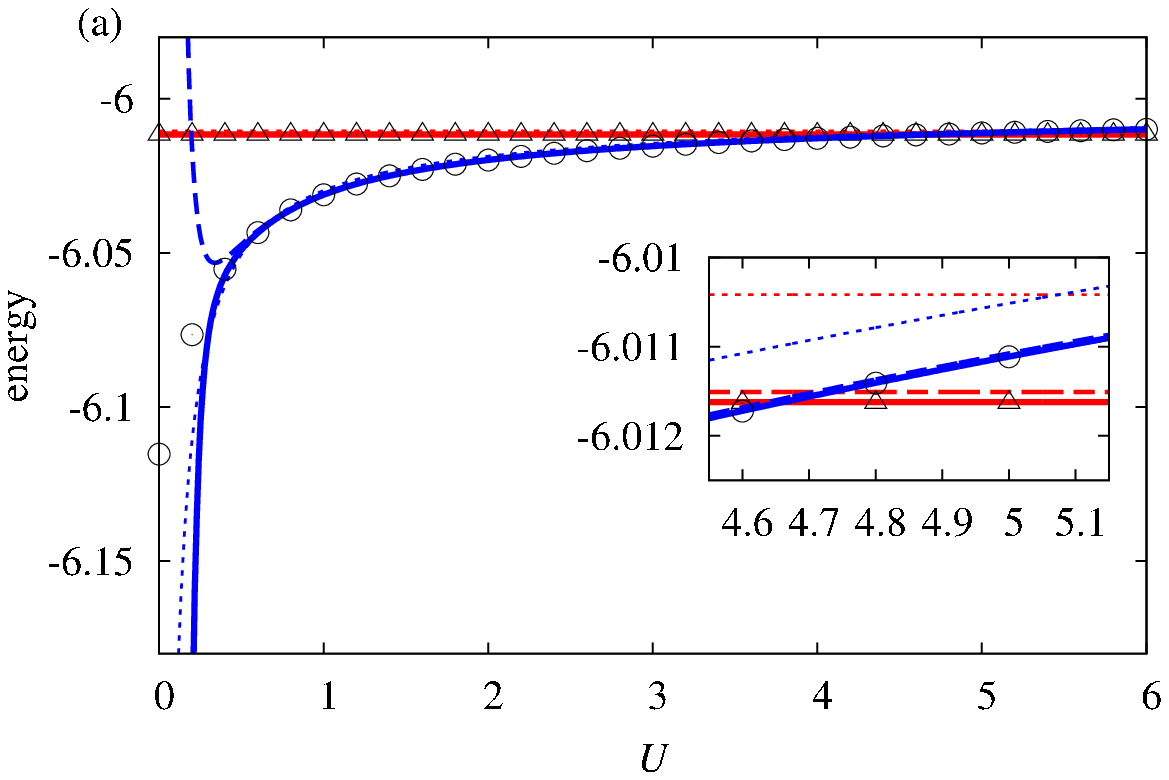}\\
\includegraphics[clip=on,width=78mm,angle=0]{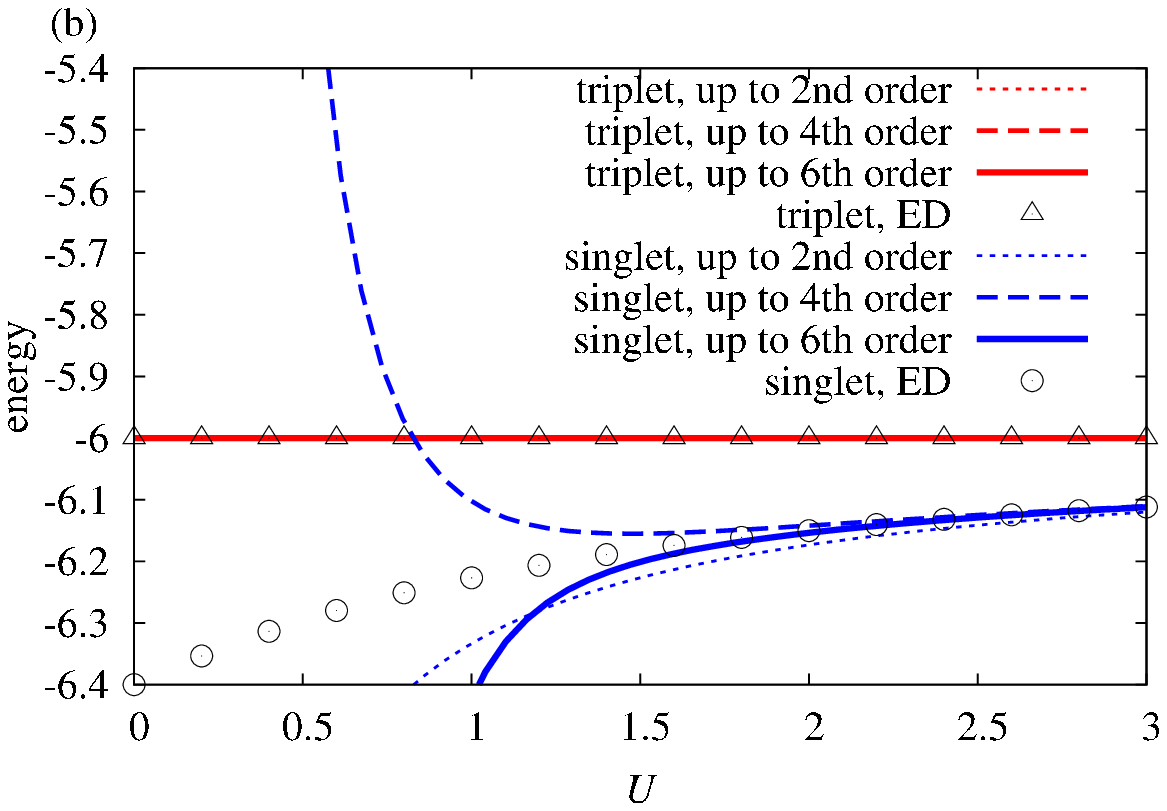}\\
\includegraphics[clip=on,width=78mm,angle=0]{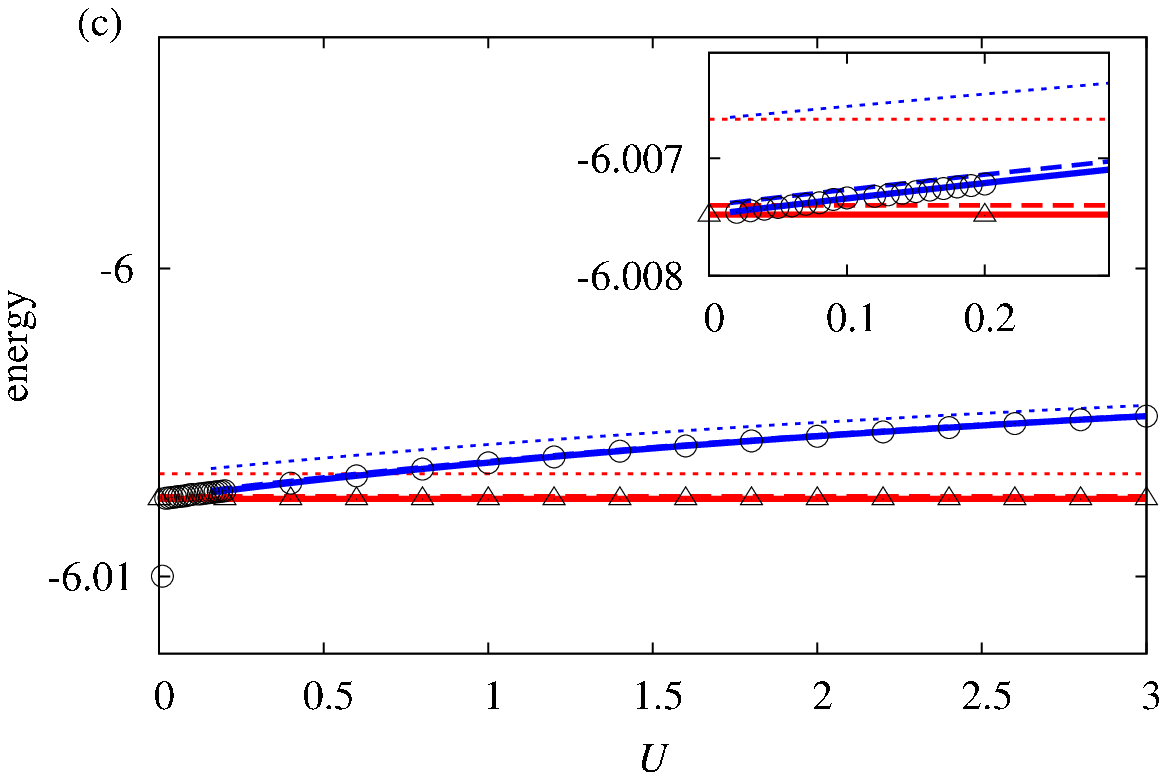} 
\caption{(Color online)
Energies of low-lying states (triplet -- red, singlet -- blue) as a function of on-site repulsion $U$  
(perturbation theory up to sixth order and exact-diagonalization data) 
for $n=2$ electrons on the ladder of ${\cal{N}}=2$ cells (open boundary conditions). 
(a)
$t_2=3$, $t_{11}=0.85$, $t_{12}=0.95$, $t_{21}=1$, $t_{22}=1.2$ (general deformation).
(b)
$t_2=3$, $t_{11}=t_{22}=1.1$, $t_{12}=t_{21}=0.9$ (symmetric deformation).
(c)
$t_2=3$, $t_{11}=t_{21}=1.1$, $t_{12}=t_{22}=0.9$ (semi-symmetric deformation);
exact diagonalization yields $U_c \approx  0.015$,
whereas the perturbation-theory prediction is $U_c^{(6)}=0$.}
\label{f03}
\end{center}
\end{figure}

We begin with the case of $n=2$ electrons on the ladder of ${\cal{N}}=2$ cells with open boundary conditions imposed.
Perturbation-theory calculations for the energies of the triplet state and the singlet state can be easily 
obtained by symbolic computation up to the sixth order:
\begin{eqnarray}
\label{505}
E_t=-2t_2    +E_t^{(2)}+E_t^{(4)}+E_t^{(6)}+\ldots,
\nonumber\\
E_s(U)=-2t_2 +E_s^{(2)}(U)+E_s^{(4)}(U)+E_s^{(6)}(U)+\ldots .
\end{eqnarray}
Here the second-order corrections are as follows:
\begin{eqnarray}
\label{506}
E_t^{(2)}=-\frac{t_l^2\delta_l^2+t_f^2\delta_f^2}{t_2},
\nonumber\\
E_s^{(2)}(U)
=
-\frac{(t_l-t_f)^2}{t_2}
-2\frac{t_l^2\delta_l^2+t_f^2\delta_f^2}{2t_2+U}
-\frac{8(t_l-t_f)^2}{U}.
\end{eqnarray}
The explicit lengthy expressions for the higher-order corrections are given in Appendix~D.
Typical dependences of low-lying energies on $U$ are shown in
Figs.~\ref{f03}(a), \ref{f03}(b), and \ref{f03}(c)
for a particular general deformation, a symmetric deformation, and a semi-symmetric deformation, respectively.

The conclusions obtained from the formulas and plots (Fig.~\ref{f03}) of the singlet and
triplet energies are as follows:
In the small-$U$ limit the perturbation theory may fail, cf. Figs.~\ref{f03}(a) and \ref{f03}(b).
The reason for this has been mentioned above already: 
In the small-$U$ limit some relevant excited states approach the ground-state manifold.
The deviation from the ideal flat-band geometry
leads to more drastic effects and
also to a larger diversity in the energy dependence on $U$ than for the diamond chain considered in the previous section.
The behavior of $E_t$ and $E_s(U)$ shown in Fig.~\ref{f03}(a) for the general
case qualitatively 
resembles that for the diamond chain (cf. Fig.~\ref{f08} in Appendix~C).
On the other hand, 
the symmetric and semi-symmetric cases are totally unlike.
Namely,
as long as the perturbation theory converges, for the symmetric deformation, case (i), the singlet energy 
(circles and blue curves) 
is always lower than the triplet energy 
(triangles and red curves), $E_s<E_t=-2t_2$, see Fig.~\ref{f03}(b).
Note that all exact-diagonalization data also yield
$E_s<E_t$ for the symmetric case. 
For the semi-symmetric case the triplet energy becomes the lowest one,
$E_t<E_s(U)$,
if $U$ exceeds a very small critical value $U_c$,
see Fig.~\ref{f03}(c).
[For the case shown in Fig.~\ref{f03}(c) 
exact diagonalization gives $U_c\approx 0.015$
and 
the perturbation-theory result is $U_c^{(6)}=0$.]
That means,
ferromagnetism does not appear at all  for the symmetric deformation, 
whereas for the semi-symmetric case only a very small $U$ is required to promote its appearance.
Next important difference in comparison to the diamond-chain case is related to the energy scale
(compare Figs.~\ref{f03} and \ref{f08}):
The splitting of triplet and singlet for the ladder occurs already in the second order 
(and only in the fourth order for the diamond chain).
This can be traced back to the difference in lattice geometries.
Thus, for the ladder  the second-order  perturbation theory already provides useful results.

The above described features of the energy dependences on $U$ 
can be understood by a more detailed analysis of the perturbation-theory
treatment, see Appendices~A and B.
For that we consider the action of the perturbation ${\sf V}$ on the triplet and singlet
states, i.e.,
${\sf V}\vert t,\pm 1\rangle$, ${\sf V}\vert t,0\rangle$, and ${\sf V}\vert s\rangle$.
The results depend on the symmetry of the imposed deformation. 
Thus, 
for the symmetric case 
${\sf V}\vert t\rangle=0$, 
but 
${\sf V}\vert s\rangle\propto (l^\dagger_{a,\uparrow}l^\dagger_{a,\downarrow}+l^\dagger_{b,\uparrow}l^\dagger_{b,\downarrow})\vert {\rm{vac}}\rangle$.
As a consequence,
the unperturbed triplet energy $-2t_2$ remains unchanged after switching on ${\sf
V}$,
whereas the unperturbed singlet energy $-2t_2$ decreases after switching on ${\sf V}$
and ferromagnetism cannot arise.
Moreover,
the state ${\sf V}\vert s\rangle$ overlaps with ``dangerous'' excited states of ${\sf{H}}_0$
(which contain $l^\dagger_{a,\uparrow}l^\dagger_{a,\downarrow}$, $l^\dagger_{b,\uparrow}l^\dagger_{b,\downarrow}$
and have the energy $-2t_2+U$ for $U\to 0$)
leading to the failure of the perturbation theory in the small-$U$ limit.
On the other hand, for the semi-symmetric case 
${\sf{V}}\vert t\rangle$
contains 
$c_{m,1,\sigma}^\dagger c_{m,2,\sigma}^\dagger\vert{\rm{vac}}\rangle$
or 
$(c_{m,1,\uparrow}^\dagger c_{m,2,\downarrow}^\dagger + c_{m,1,\downarrow}^\dagger
c_{m,2,\uparrow}^\dagger)\vert{\rm{vac}}\rangle$,
whereas
${\sf{V}}\vert s\rangle\propto (c_{m,1,\uparrow}^\dagger c_{m,1,\downarrow}^\dagger - c_{m,2,\uparrow}^\dagger c_{m,2,\downarrow}^\dagger)\vert{\rm{vac}}\rangle$.
Since the state 
$(c_{m,1,\uparrow}^\dagger c_{m,1,\downarrow}^\dagger - c_{m,2,\uparrow}^\dagger c_{m,2,\downarrow}^\dagger)\vert{\rm{vac}}\rangle$
is orthogonal to the dangerous excited states of ${\sf{H}}_0$,
the perturbation theory does not fail in the small-$U$ limit.
Moreover,
the states ${\sf{V}}\vert t\rangle$ and ${\sf{V}}\vert s\rangle$ have the same overlap integral with the excited states of ${\sf{H}}_0$ with the energies 0 and $U$, respectively.
Therefore, the decrease of the triplet energy exceeds the decrease of the singlet energy
instantaneously as $U>0$, 
i.e., ferromagnetism appears for infinitesimally small positive $U$.

In second order the perturbation theory yields a compact formula 
for the critical value of on-site repulsion $U_c$.
Using Eq.~(\ref{506}) we get
\begin{eqnarray}
\label{507}
\frac{U_c^{(2)}}{t_2}
=
\frac{5\vert t_l-t_f\vert +\sqrt{9(t_l-t_f)^2+16(t_l^2\delta_l^2+t_f^2\delta_f^2)}}{-(t_l-t_f)^2+t_l^2\delta_l^2+t_f^2\delta_f^2}
\nonumber\\
\times\vert t_l-t_f\vert. \qquad
\end{eqnarray}
Obviously,
for symmetric deformations, 
when $t_l\ne t_f$ and $\delta_l=\delta_f=0$, 
Eq.~(\ref{507}) gives for $U_c^{(2)}=-8t_2 <0$,
that is consistent with  the absence of  ferromagnetism in this case.
It is also obvious, that formula (\ref{507})
yields $U^{(2)}_c=0$ for $t_l=t_f$, 
i.e., for $t_{11}-t_{12}-t_{21}+t_{22}=0$.
That criterion,  $t_l=t_f$,  holds 
for semi-symmetric deformations, where in addition also $\delta_l=\delta_f$
is valid.
However, in higher-order perturbation theory as well as in exact diagonalization
we find that the constraint  $t_l=t_f$ does not imply $U_c=0$, rather $U_c$
may become large for the general case $\delta_l\ne \delta_f$, if
$\delta_l$ or $\delta_f$ become of the order of unity, see Fig.~\ref{f06}.

\begin{figure}%[h]
\begin{center}
\includegraphics[clip=on,width=72mm,angle=0]{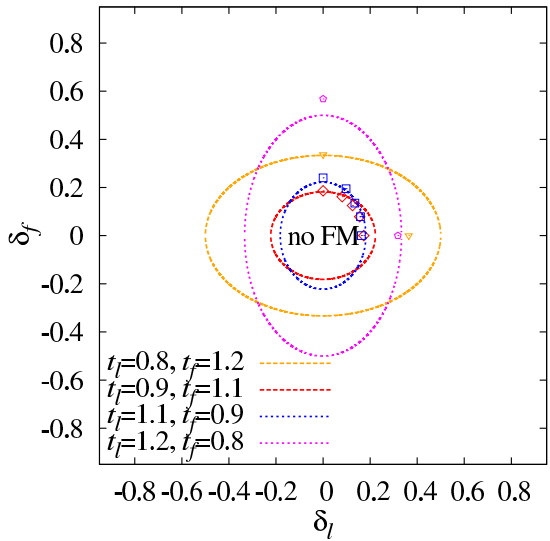}
\caption{(Color online)
There is no ferromagnetic ground states for the Hubbard ladder
in the region around the origin of the plane $\delta_l$--$\delta_f$
[$t_{11}-t_{12}-t_{21}+t_{22}=2(t_l-t_f)\ne 0$].
Analytical predictions based on the second-order perturbation-theory calculations (\ref{507})
(lines)
are compared with exact-diagonalization data for $N=16$, $t_2=3$
(symbols)
for several values of $t_l$ and $t_f$, $t_l+t_f=2$.}
\label{f04}
\end{center}
\end{figure}

Supposing that the energies behave smoothly as changing deformations, 
we can expect that there is a finite parameter region in the vicinity of
the symmetric case without ground-state ferromagnetism.
Indeed, for $t_l\ne t_f$  the second-order formula (\ref{507}) leads to an elliptic shape
in the  $\delta_l$--$\delta_f$ plane given by
\begin{eqnarray}
\label{508}
\left(\frac{t_l}{t_l-t_f}\delta_l\right)^2+\left(\frac{t_f}{t_l-t_f}\delta_f\right)^2=1.
\end{eqnarray}
We illustrate this behavior in Fig.~\ref{f04}, where we also show a few
points obtained by exact diagonalization which are in qualitative agreement
with the predictions from Eq.~(\ref{508}).
It is worthwhile to remark 
that Eq.~(\ref{508}) remains unaltered if interchanging 
$t_l \leftrightarrow t_f$ and $\delta_l \leftrightarrow \delta_f$
(this symmetry is also evident in Fig.~\ref{f04}).
However,
exact-diagonalization data shown by symbols in Fig.~\ref{f04} do not show this symmetry present in the second-order results, 
i.e., it is not generally  present in the model,
cf., e.g., Eq.~(\ref{502}).

\subsection{Three (four) electrons and three (four) cells}

Let us discuss briefly the perturbation theory for larger clusters.
In the case of three electrons on the ladder of three cells we face a $2^3$-fold degenerate ground
state,
which consists of the quadruplet $\vert q\rangle$ (total spin is 3/2) and two doublets $\vert d1\rangle$ and $\vert d2\rangle$ (total spin is 1/2).
We are interested in the energies $E_q$, $E_{d1}$, and $E_{d2}$.
In Appendix~D,
we provide explicit expressions for these energies 
\begin{eqnarray}
\label{509}
E_q=-3t_2+E_q^{(2)}+E_q^{(4)}+\ldots,
\nonumber\\
E_{di}(U)=-3t_2+E_{di}^{(2)}(U)+E_{di}^{(4)}(U)+\ldots,
\nonumber\\
i=1,2.
\end{eqnarray}
In the case of four electrons on the ladder of four cells we face a $2^4$-fold degenerate ground state,
which consists of the quintuplet $\vert Q\rangle$ (total spin is 2),
three triplets $\vert t1\rangle$, $\vert t2\rangle$, $\vert t3\rangle$ (total spin is 1),
and two singlets $\vert s1\rangle$, $\vert s2\rangle$ (total spin is 0).
In Appendix~D,
we provide explicit expressions for their energies 
\begin{eqnarray}
\label{510}
E_Q=-4t_2+E_Q^{(2)}+E_Q^{(4)}+\ldots,
\nonumber\\
E_{ti}(U)=-4t_2+E_{ti}^{(2)}(U) +E_{ti}^{(4)}(U)+\ldots,
\nonumber\\
i=1,2,3,
\nonumber\\
E_{sj}(U)=-4t_2+E_{sj}^{(2)}(U) +E_{sj}^{(4)}(U)+\ldots,
\nonumber\\
j=1,2.
\end{eqnarray}
We report corresponding results for the energies up to the fourth order along with exact-diagonalization data 
for the general, symmetric, and semi-symmetric deformations
for $n={\cal{N}}=3$ and $n={\cal{N}}=4$ in Appendix~D.
The main features of these results resemble strongly the ones discussed in the previous subsection for $n={\cal{N}}=2$.
Therefore, the main conclusions obtained from those data for the energies of
larger cells are
consistent with those discussed in Sec.~\ref{ladder_two_cells} for two
cells. 
Most remarkably,
within the second-order perturbation theory, 
the critical value $U_c^{(2)}$ for the three-cell and four-cell clusters 
coincide with $U_c^{(2)}$ for the two-cell cluster, 
i.e., it is given by Eq.~(\ref{507}).

Let us finally mention that within the perturbation theory for ${\cal{N}}=4$ cells  
the fully polarized ferromagnetic state (it is a quintuplet for ${\cal{N}}=4$) 
is in competition with triplet and singlet states. 
We find, 
cf. Fig.~\ref{f11},
that either a singlet or the ferromagnetic quintuplet is the ground state.
This finding, 
that the fully polarized ferromagnetic state competes with a nonmagnetic singlet state 
(but not with partially polarized states) 
is supported by exact-diagonalization data obtained for systems with an even number of cells ${\cal{N}}>4$.        

\subsection{Phase diagram}

In this subsection we collect analytical and numerical findings
to construct the ground-state phase diagrams of the Hubbard ladder.
According to Eq.~(\ref{203}),
there are three parameters which characterize the ladder,
i.e., $t_l$ and $t_f$ with $t_l+t_f=2t<t_2$, $\delta_l$, and $\delta_f$.
We set $t_2=3$, $t_l+t_f=2$.
After fixing  $t_l$ and $t_f$ we are left with two free parameters $\delta_l$ and $\delta_f$.
We consider the first quadrant of positive $\delta_l$ and $\delta_f$ in the $\delta_l$--$\delta_f$ plane.  
We move through the quadrant by straight lines 
in the horizontal direction ($\delta_f$ is fixed, $\delta_l$ varies),
in the vertical direction ($\delta_l$ is fixed, $\delta_f$ varies),
as well as along the diagonal $\delta_l=\delta_f=\delta$.
Certainly perturbation-theory results are reasonable only for small deviations from the ideal flat-band geometry.
However, there are no such restrictions for exact-diagonalization data.

\begin{figure}%[h]
\begin{center}
\includegraphics[clip=on,width=62.5mm,angle=0]{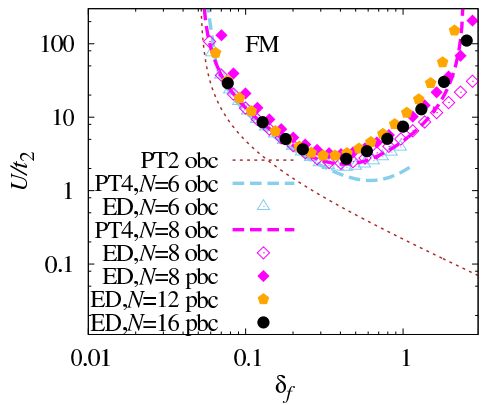}
\caption{(Color online)
Phase diagram in the quarter plane $\delta_f$ -- $U/t_2$ 
for the ladder with $t_2=3$, $t_l=1.025$, $t_f=0.975$, and $\delta_l=0$
obtained by perturbation-theory calculations and by exact diagonalization 
for $N=6,\;8,\;12,\;16$ with open and/or periodic boundary conditions.}
\label{f05}
\end{center}
\end{figure}

We begin with a quite general case assuming $t_l=1.025$, $t_f=0.975$ and $\delta_l=0$.
The dependence of $U_c$ on $\delta_f$ is reported in Fig.~\ref{f05}.
The ground state is ferromagnetic above the curves $U_c(\delta_f)$;
this region is denoted as FM. 
In this case,
the dependence of $U_c$ on the acquired bandwidth is a nonmonotonic function:
For small $\delta_f$ ferromagnetism does not appear at all 
[in agreement with Eq.~(\ref{508})];
increasing  $\delta_f$ beyond a threshold value $\delta_{f1}$ 
ferromagnetism sets in and $U_c$ decreases with growing $\delta_f$.
Second-order perturbation theory, Eq.~(\ref{508}), predicts
$\delta_{f1}\approx 0.051$,
exact diagonalization for $N=16$ yields
$\delta_{f1}\approx 0.053$.
Beyond $\delta_f\approx 0.4$ the critical repulsion  $U_c$ starts to increase with increasing of $\delta_f$.
This behavior is obtained from both the fourth-order perturbation theory and exact diagonalization
for different system sizes with open and/or periodic boundary conditions imposed.
The second-order perturbation theory gives qualitatively correct results only for small $\delta_f<0.4$.
From exact-diagonalization data for $N=16$ it is obvious 
that there is  again a threshold value $\delta_{f2}$  
(for  $N=16$ we found $\delta_{f2} \approx 3.25$)
above which no ferromagnetism appears.
Fig.~\ref{f05} illustrates a quite subtle interplay of the hopping-integral geometry and the on-site Hubbard repulsion
required for establishing of ground-state ferromagnetism.

\begin{figure}%[h]
\begin{center}
\includegraphics[clip=on,width=62.5mm,angle=0]{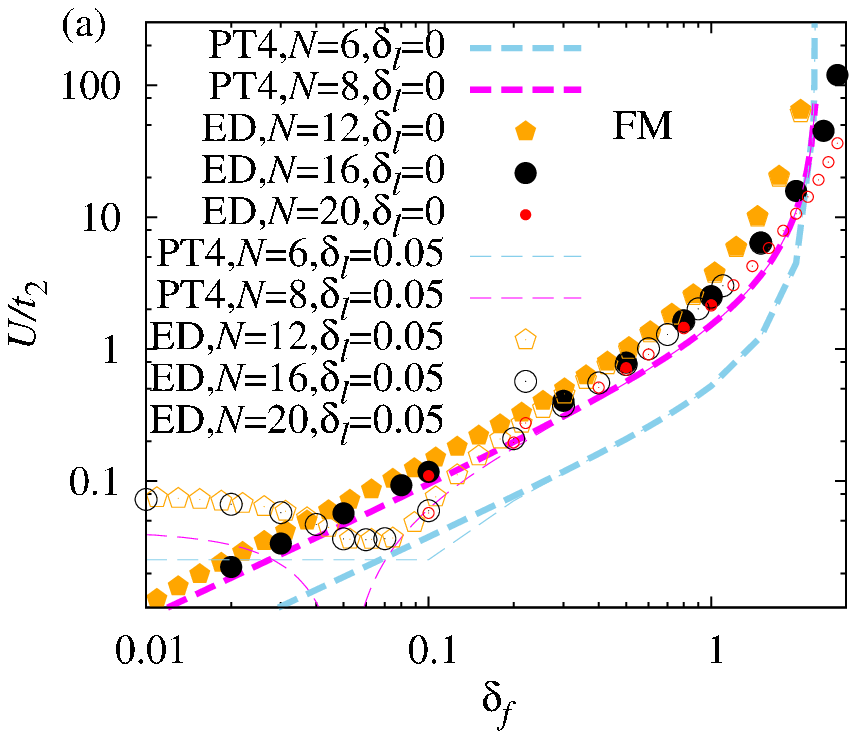}\\
\includegraphics[clip=on,width=62.5mm,angle=0]{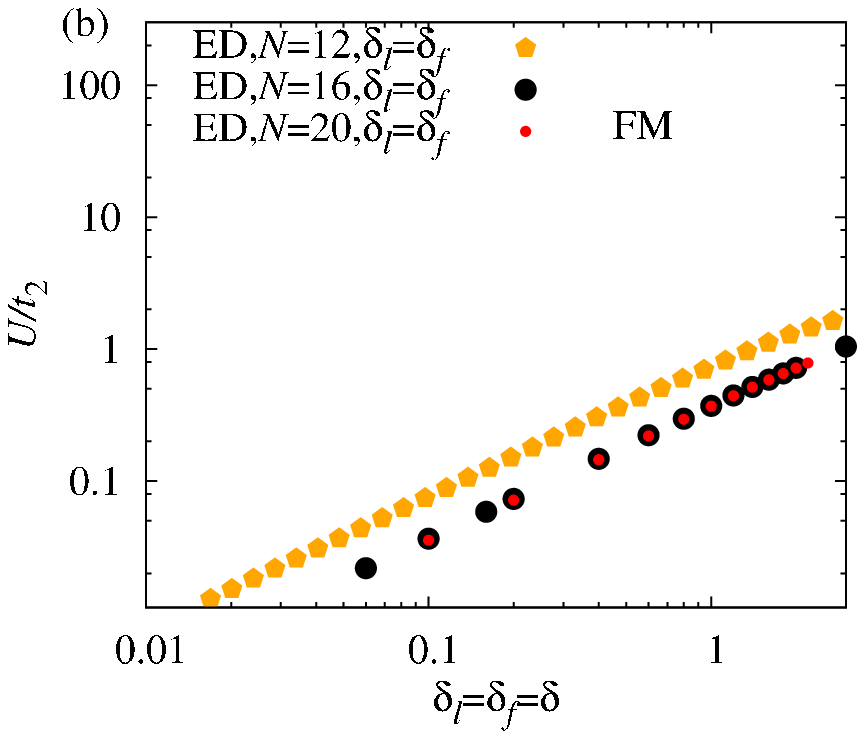}\\
\includegraphics[clip=on,width=62.5mm,angle=0]{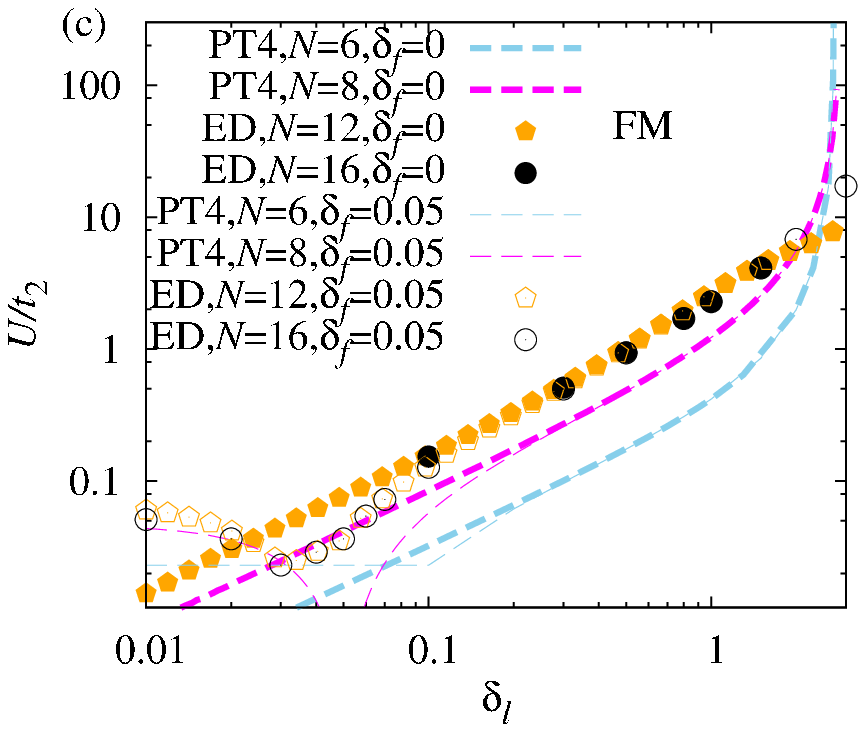}
\caption{(Color online)
Phase diagram in the quarter plane $\delta$ -- $U/t_2$
for the ladder with $t_2=3$, $t_l=t_f=1$ 
obtained by fourth-order perturbation theory 
and by exact diagonalization 
(note that second-order perturbation theory yields $U^{(2)}_c=0$).
(a)
$\delta_l=0,\;0.05$.
(b)
$\delta_l=\delta_f=\delta$;
fourth-order perturbation-theory calculations yield zero value for $U_c$.
(c)
$\delta_f=0,\;0.05$.}
\label{f06}
\end{center}
\end{figure}

Next we pass to the case $t_l=t_f=1$.
The dependences of $U_c$ on $\delta_f$, on $\delta_l=\delta_f=\delta$, and on $\delta_l$
are reported in panels (a), (b), and (c) in Fig.~\ref{f06}, respectively. 
The ground state is ferromagnetic above the curves $U_c(\delta)$;
this region is denoted as FM.
We recall that in the case $t_l=t_f$ from Eq.~(\ref{507}) we get $U^{(2)}_c=0$;
nonzero values of $U_c$ come only from higher-order (in fact, fourth-order) calculations.
Furthermore,
for the semi-symmetric deformation, 
i.e., $\delta_l=\delta_f=\delta$,
the perturbation theory yields $U_c^{(4)}=0$. 
Obviously, higher-order processes should lead to finite values for $U_c$,  
as it is indicated by the exact-diagonalization data shown Fig.~\ref{f06}(b).

As can be seen in Figs.~\ref{f06}(a)  and \ref{f06}(c),
analytical results which refer to the case of ${\cal{N}}=3,\,4$ cells with open boundary conditions
and
exact-diagonalization data which refer to the case of ${\cal{N}}=6,\,8,\,10$ cells 
are in a reasonable agreement. 
By contrast to the parameter situation shown in Fig.~\ref{f05}, 
in all cases presented in Fig.~\ref{f06} 
ground-state ferromagnetism can be obtained also for small deviations from the flat-band geometry 
(controlled by $\delta_f$ and/or $\delta_l$).
Comparing the exact-diagonalization data for different system sizes $N$ 
we observe that the finite-size effects remain small, 
thus  the discussed phenomenon should be present for thermodynamically large systems, too.

It is in order to mention a special finite-size effect 
that may appear  for large values of $\delta_l$ and/or  $\delta_f$.
In this limit, the dominating hopping parameters may correspond to geometries 
which do not fit to the initial ladder structure.
Thus, for $t_{11}=1+\delta_l$, $t_{22}=1-\delta_l$ and small $\delta_f$, 
in the limit of $\delta_l\to\infty$ the legs of the ladder form two almost decoupled chains. 
Such a finite simple Hubbard chain  at quarter filling with an odd number of electrons 
(i.e., a chain of 6 or 10 sites with 3 or 5 electrons) 
has a ferromagnetic ground state.
Therefore, the limit of large deviations, shown for completeness  in our
figures, goes beyond the primary focus of discussing the dispersion-driven
ferromagnetism in systems with ladder geometry.

\section{Bilayer}
\label{sec6}
\setcounter{equation}{0}

As mentioned already, the mechanism leading  to the emergence of ferromagnetism
driven by kinetic energy is not restricted to dimension $D=1$.
To illustrate this, we consider the two-dimensional counterpart of the
Hubbard ladder, namely the frustrated bilayer,
see Fig.~\ref{f01}.
From the technical point of view, the two-dimensional model is more
challenging, since the smallest cluster appropriate for perturbation theory  and imaging the basic geometry of the
bilayer is built
by five cells (a central cell with four neighboring cells). 
Furthermore, in contrast to the ladder for the exact diagonalization we do not have a
sequence of finite lattices of $N=12,16,20$ sites in $D=2$. The smallest finite
bilayer lattice 
with periodic boundary conditions has $N=16$ sites.
Hence, we cannot provide a detailed discussion of the bilayer model, rather
we will demonstrate for a particular parameter set that the mechanism of kinetic-energy-driven 
ferromagnetism also holds in $D=2$.

In analogy to the ladder, 
for the bilayer one of the two one-electron bands is flat if
$t_{11}=t_{12}=t_{21}=t_{22}=t$ or $t_l=t_f=t$, $\delta_l=\delta_f=0$ 
and it becomes the lowest one if
$4t<t_2$.
Within fourth-order perturbation theory we are able to calculate 
the energies of the fully polarized sextuplet (total spin 5/2)
 and of the quadruplets (total spin 3/2),
\begin{eqnarray}
E_S=-5t_2+E_S^{(2)}+E_S^{(4)}+\ldots,
\nonumber\\
E_{qi}(U)=-5t_2+E_{qi}^{(2)}(U)+E_{qi}^{(4)}(U)+\ldots,
\nonumber\\
i=1,2,3,4,
\end{eqnarray}
see Appendix~E.
Hence, our perturbation-theory treatment remains incomplete, 
since we cannot compare with the energies of the five doublets with total spin 1/2. 
On the other hand, 
the comparison with the exact-diagonalization data for the five-cell cluster, 
where the doublet states are taken into account, 
yields an excellent agreement between both approaches. 
That is because for this cluster the level crossing between the sextuplet and the lowest quadruplet 
takes place at the same $U$ as for the crossing of  sextuplet and the lowest doublet.

\begin{figure}%[h]
\begin{center}
\includegraphics[clip=on,width=62.5mm,angle=0]{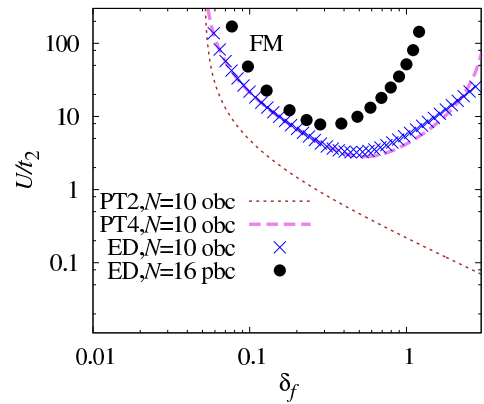}
\caption{(Color online)
Phase diagram in the quarter plane $\delta_f$ -- $U/t_2$ 
for the bilayer with $t_2=5$, $t_l=1.025$, $t_f=0.975$, and $\delta_l=0$
obtained by perturbation-theory calculations and by exact diagonalization 
for $N=10,\,16$ with open and periodic boundary conditions.}
\label{f07}
\end{center}
\end{figure}

As a first (remarkable) outcome we find,
that the second-order result $U_c^{(2)}$ again is given by Eq.~(\ref{507}).
We show numerical data for the critical repulsion $U_c$  for the set of parameters
$t_2=5$, $t_l=1.025$, $t_f=0.975$, and $\delta_l=0$ 
in the ground-state phase diagram
presented  in Fig.~\ref{f07} (cf. the corresponding phase diagram for the ladder shown
Fig.~\ref{f05}).

Basically  the same features as for the corresponding ladder are also found
for the phase diagram of the bilayer.
However, it is obvious that $U_c$ for the finite lattice of $N=16$ sites with periodic boundary conditions 
is noticeably above perturbation-theory results and the exact-diagonalization results for $N=10$ sites.      
We argue that the finite system of $N=10$ sites with open boundary conditions
is only a very rough model of thermodynamically large bilayer,
since only one (among five) vertical bond has the same environment as in infinite lattice.
The finite system of $N=16$ sites with periodic boundary conditions is free of this shortcoming.

\section{Conclusions}
\label{sec7}
\setcounter{equation}{0}

We have used perturbation theory as well as exact diagonalization of
finite systems 
to examine the kinetic-energy-driven emergence of ferromagnetic ground states 
in Hubbard models with a half-filled lowest-energy flat band for lattices
which do not obey the connectivity condition (isolated trapping cells). 
Generally speaking, 
if 
(i) the flat band acquires a small dispersion this way allowing to the previously localized electrons to correlate 
and
(ii) the on-site Hubbard repulsion $U$ is sufficiently strong
the ground state becomes ferromagnetic.
However, 
the relation between the required $U_c$ and the acquired bandwidth might be quite
intricate.
Thus,
for some deformation geometries ferromagnetism does not appear at all, 
for others it appears already for small $U$;
in some cases $U_c$ is an increasing function of the deformation strength,
whereas in others it becomes nonmonotonic.
The mechanism leading to kinetic-energy-driven emergence of ferromagnetism
is studied in detail for one-dimensional systems with isolated trapping cells.
However, as it is demonstrated for a specific two-dimensional system this mechanism 
works in higher dimensions as well.
Although our analysis refers to finite systems,
the observed finite-size behavior indicates convincingly that such a scenario should survive in the thermodynamic
limit, too.
Thus our main conclusion is 
that the described phenomenon is a quite general way of establishing ground-state ferromagnetism 
in the repulsive Hubbard model at low electron densities  
around the flat-band limit.

Furthermore,
for special examples, 
the diamond chain, the ladder as well as the bilayer, 
we have obtained simple analytical formulas, cf. Eq.~(\ref{403}) and Eq.~(\ref{507}),
which amazingly well estimate the region of ground-state ferromagnetism.
From the technical point of view,
we have elaborated computer-adapted scheme for analytical perturbation-theory calculations up to the sixth order.

Finally,
it is in order to notice
that experimental searches for Mielke-Tasaki flat-band ferromagnetism remain an ambitious goal of numerous experimental studies,
see, e.g., Refs.~\onlinecite{ijmpb,experiments}.
Our findings offer new perspectives for investigating solid-state realization of flat-band
ferromagnetism, since the emergence of ferromagnetism in systems with isolated
trapping cells does not require  fine tuning of
parameters, rather it can be found in a quite wide parameter region.

\section*{Acknowledgments}

The present study was supported by the DFG (project RI615/21-1).
O.~D. acknowledges the kind hospitality of the University of Magdeburg in April-May and October-December of 2015.
O.~D. would like to thank the Abdus Salam International Centre for Theoretical Physics (Trieste, Italy)
for partial support of this study through the Senior Associate award.

\onecolumngrid

\section*{Appendix A: Perturbation-theory formulas for the ground-state energy up to the sixth order}
\renewcommand{\theequation}{A\arabic{equation}}
\setcounter{equation}{0}

In this appendix,
we present the perturbation-theory formulas up to the sixth order, which are used in our study.
Although these formulas can be found in Ref.~\onlinecite{klein},
we show them here for the reader's convenience and the self-consistency of the paper.

First we split the ${\cal{N}}$-cell Hamiltonian of the model 
$H$ into the main part ${\sf{H}}_0$ and the perturbation ${\sf{V}}$,
i.e., $H={\sf{H}}_0+{\sf{V}}$.
We consider the subspace of $n={\cal{N}}$ electrons. 
All eigenstates $\vert\alpha\rangle$ and their energies 
${\sf{E}}_{\alpha}$ of the unperturbed Hamiltonian ${\sf{H}}_0$ are known.
We consider the ground state $\vert{\rm{GS}}\rangle$ of the unperturbed Hamiltonian ${\sf{H}}_0$,
which is $2^n$-fold degenerate 
(each cell can be occupied either by up- or down-spin electron). 
We denote the ground-state energy by ${\sf{E}}_{{\rm{GS}}}$.
Moreover, we have $\langle{\rm{GS}}\vert{\sf{V}}\vert{\rm{GS}}\rangle=0$.
Since the ground states are degenerate, the choice of the ground states
requires some consideration.
From Ref.~\onlinecite{oleg-johannes} we know that the effective Hamiltonian to
describe the low-energy degrees of freedom is a Heisenberg Hamiltonian. Hence, we choose the set of ground
states as a corresponding set of eigenstates of the Heisenberg model that
way also
implying the required  SU(2) symmetry as well as the spatial symmetry of the
clusters used for the perturbation theory (for details see  Appendix~B). 
The lowest-order perturbation-theory corrections to the ground-state energy ${\sf{E}}_{{\rm{GS}}}$ are as follows:
\begin{eqnarray}
\label{a01}
E^{(2)}_{{\rm{GS}}}=\sideset{}{'}\sum_{\alpha}
\frac{\langle{\rm{GS}}\vert{\sf{V}}\vert\alpha\rangle\langle\alpha\vert{\sf{V}}\vert{\rm{GS}}\rangle}
{{\sf{E}}_{{\rm{GS}}}-{\sf{E}}_{\alpha}},
\nonumber\\
E^{(3)}_{{\rm{GS}}}=\sideset{}{'}\sum_{\alpha} \sideset{}{'}\sum_{\beta}
\frac{\langle{\rm{GS}}\vert{\sf{V}}\vert\alpha\rangle\langle\alpha\vert{\sf{V}}\vert\beta\rangle
\langle\beta\vert{\sf{V}}\vert{\rm{GS}}\rangle}
{\left({\sf{E}}_{{\rm{GS}}}-{\sf{E}}_{\alpha}\right)\left({\sf{E}}_{{\rm{GS}}}-{\sf{E}}_{\beta}\right)},
\nonumber\\
E^{(4)}_{{\rm{GS}}}=\sideset{}{'}\sum_{\alpha} \sideset{}{'}\sum_{\beta} \sideset{}{'}\sum_{\gamma}
\frac{\langle{\rm{GS}}\vert{\sf{V}}\vert\alpha\rangle\langle\alpha\vert{\sf{V}}\vert\beta\rangle
\langle\beta\vert{\sf{V}}\vert\gamma\rangle\langle\gamma\vert{\sf{V}}\vert{\rm{GS}}\rangle}
{\left({\sf{E}}_{{\rm{GS}}}-{\sf{E}}_{\alpha}\right)\left({\sf{E}}_{{\rm{GS}}}-{\sf{E}}_{\beta}\right)
\left({\sf{E}}_{{\rm{GS}}}-{\sf{E}}_{\gamma}\right)}
-\sideset{}{'}\sum_{\alpha} \sideset{}{'}\sum_{\beta}
\frac{\langle{\rm{GS}}\vert{\sf{V}}\vert\alpha\rangle\langle\alpha\vert{\sf{V}}\vert{\rm{GS}}\rangle
\langle{\rm{GS}}\vert{\sf{V}}\vert\beta\rangle\langle\beta\vert{\sf{V}}\vert{\rm{GS}}\rangle}
{\left({\sf{E}}_{{\rm{GS}}}-{\sf{E}}_{\alpha}\right)^2\left({\sf{E}}_{{\rm{GS}}}-{\sf{E}}_{\beta}\right)},
\nonumber\\
E^{(5)}_{{\rm{GS}}}=(1,1,1,1)
+\frac{1}{2}(2,1,0,1)+\frac{1}{2}(1,2,0,1)+\frac{1}{2}(1,1,0,2)
+\frac{1}{2}(2,0,1,1)+\frac{1}{2}(1,0,2,1)+\frac{1}{2}(1,0,1,2),
\nonumber\\
E^{(6)}_{{\rm{GS}}}
=
(1,1,1,1,1)
\nonumber\\
+\frac{1}{2}(2,1,1,0,1)+\frac{1}{2}(1,2,1,0,1)+\frac{1}{2}(1,1,2,0,1)+\frac{1}{2}(1,1,1,0,2)
\nonumber\\
+\frac{1}{2}(2,1,0,1,1)+\frac{1}{2}(1,2,0,1,1)+\frac{1}{2}(1,1,0,2,1)+\frac{1}{2}(1,1,0,1,2)
\nonumber\\
+\frac{1}{2}(2,0,1,1,1)+\frac{1}{2}(1,0,2,1,1)+\frac{1}{2}(1,0,1,2,1)+\frac{1}{2}(1,0,1,1,2)
\nonumber\\
+\frac{1}{2}(3,0,1,0,1)+\frac{3}{8}(2,0,2,0,1)+\frac{1}{4}(2,0,1,0,2)+\frac{3}{8}(1,0,2,0,2)+\frac{1}{2}(1,0,1,0,3);
\end{eqnarray}
here the superscript `prime' means 
that the sum extends over all states of the unperturbed Hamiltonian ${\sf{H}}_0$ except the ground states.
Moreover, 
we have introduced shorthand notations\cite{klein}
\begin{eqnarray}
\label{a02}
(k_1,k_2,\ldots,k_n)
=\langle{\rm{GS}}\vert {\sf{V}} R^{(k_1)} {\sf{V}} R^{(k_2)} {\sf{V}}\ldots {\sf{V}} R^{(k_n)} {\sf{V}} \vert{\rm{GS}}\rangle,
\nonumber\\
R^{(k)}=
\left\{
\begin{array}{ll}
-\vert{\rm{GS}}\rangle \langle{\rm{GS}}\vert,                                                                                 & k=0,\\
\left(\sideset{}{'}\sum_{\alpha}\frac{\vert\alpha\rangle\langle\alpha\vert}{{\sf{E}}_{{\rm{GS}}}-{\sf{E}}_{\alpha}}\right)^k, & k>0
\end{array}
\right.
\end{eqnarray}
(again the superscript `prime' means 
that the sum extends over all states of the unperturbed Hamiltonian ${\sf{H}}_0$ except the ground state)
in the formulas for $E^{(5)}_{{\rm{GS}}}$ and $E^{(6)}_{{\rm{GS}}}$.

In the present study we are able to calculate the sixth-order corrections for
the ${\cal{N}}=2$-cell cases, but 
fourth-order corrections for the cases of ${\cal{N}}=3$, ${\cal{N}}=4$, and ${\cal{N}}=5$ cells.

\section*{Appendix B: Ground states of the unperturbed Hamiltonian}
\renewcommand{\theequation}{B\arabic{equation}}
\setcounter{equation}{0}
The energy of the $2^n$-fold degenerate (see Appendix~A) unperturbed ground states is ${\sf{E}}_{{\rm{GS}}}=-nt_2$.
Before applying perturbation-theory formulas of Appendix~A
we have to construct within $2^n$-fold degenerate ground states 
the ``correct'' $2^n$ linear combinations 
being SU(2) symmetric eigenstates of the corresponding Heisenberg model of the perturbation-theory clusters.
The energy of all components of a SU(2) multiplet is the same 
(i.e., are not splitted by the perturbation $\sf{V}$).
However, the energies of different multiplets may become different after switching
on perturbation, where at least second-order theory is required, since
$\langle{\rm{GS}}\vert{\sf{V}}\vert{\rm{GS}}\rangle=0$.
Thus, the number of different energies obtained by  perturbation theory cannot 
exceed $2,\,3,\,6,\,10$ for the case of ${\cal{N}}=2,\,3,\,4,\,5$ cells, respectively.

We begin with the case of ${\cal{N}}=2$ cells ($m=1$ and $m+1=2$ in Fig.~\ref{f01}) and $n=2$ electrons. 
``Correct'' unperturbed ground states are as follows:
\begin{eqnarray}
\label{b01}
\vert t,1\rangle=l^{\dagger}_{1,\uparrow}l^{\dagger}_{2,\uparrow}\vert 0\rangle, 
\;\;\;
\vert t,0\rangle=\frac{1}{\sqrt{2}}\left(l^{\dagger}_{1,\uparrow}l^{\dagger}_{2,\downarrow}+l^{\dagger}_{1,\downarrow}l^{\dagger}_{2,\uparrow}\right)\vert 0\rangle,  
\;\;\;
\vert t,-1\rangle=l^{\dagger}_{1,\downarrow}l^{\dagger}_{2,\downarrow}\vert 0\rangle,
\nonumber\\
\vert s\rangle=\frac{1}{\sqrt{2}}\left(l^{\dagger}_{1,\uparrow}l^{\dagger}_{2,\downarrow}-l^{\dagger}_{1,\downarrow}l^{\dagger}_{2,\uparrow}\right)\vert 0\rangle,
\end{eqnarray}
i.e., the three components of the triplet states $\vert t\rangle$ 
and the singlet state $\vert s\rangle$.
It is convenient to use shorthanded notations
$\vert\uparrow\uparrow\rangle=l^{\dagger}_{1,\uparrow}l^{\dagger}_{2,\uparrow}\vert 0\rangle$, 
$\vert\uparrow\downarrow\rangle=l^{\dagger}_{1,\uparrow}l^{\dagger}_{2,\downarrow}\vert 0\rangle$ etc. 
so that Eq.~(\ref{b01}) becomes
\begin{eqnarray}
\label{b02}
\vert t,1\rangle=\vert\uparrow\uparrow\rangle, 
\;\;\;
\vert t,0\rangle=\frac{1}{\sqrt{2}}\left(\vert\uparrow\downarrow\rangle + \vert\downarrow\uparrow\rangle\right),
\;\;\;
\vert t,-1\rangle=\vert\downarrow\downarrow\rangle,
\nonumber\\
\vert s\rangle=\frac{1}{\sqrt{2}}\left(\vert\uparrow\downarrow\rangle-\vert\downarrow\uparrow\rangle\right).
\end{eqnarray}

We pass to the case of ${\cal{N}}=3$ cells (open boundary conditions) and $n=3$ electrons. 
``Correct'' unperturbed ground states  are
\begin{eqnarray}
\label{b03}
\left\vert q,\frac{3}{2}\right\rangle=\vert\uparrow\uparrow\uparrow\rangle, 
\left\vert q,\frac{1}{2}\right\rangle
=\frac{1}{\sqrt{3}}\left(\vert\uparrow\uparrow\downarrow\rangle+\vert\uparrow\downarrow\uparrow\rangle+\vert\downarrow\uparrow\uparrow\rangle\right),
\left\vert q,-\frac{1}{2}\right\rangle
=\frac{1}{\sqrt{3}}\left(\vert\uparrow\downarrow\downarrow\rangle+\vert\downarrow\uparrow\downarrow\rangle+\vert\downarrow\downarrow\uparrow\rangle\right),
\left\vert q,-\frac{3}{2}\right\rangle=\vert\downarrow\downarrow\downarrow\rangle,
\nonumber\\
\left\vert d1,\frac{1}{2}\right\rangle
=  
\frac{1}{\sqrt{2}}
\left(\vert \uparrow\uparrow\downarrow\rangle - \vert\downarrow\uparrow\uparrow\rangle \right),
\qquad
\nonumber\\
\left\vert d1,-\frac{1}{2}\right\rangle
=  
\frac{1}{\sqrt{2}}
\left(\vert\uparrow\downarrow\downarrow\rangle -\vert\downarrow\downarrow\uparrow\rangle\right),
\qquad
\nonumber\\
\left\vert d2,\frac{1}{2}\right\rangle
=\frac{1}{\sqrt{6}}
\left( \vert\uparrow\uparrow\downarrow\rangle - 2\vert\uparrow\downarrow\uparrow\rangle + \vert\downarrow\uparrow\uparrow\rangle \right),
\qquad
\nonumber\\
\left\vert d2,-\frac{1}{2}\right\rangle
=\frac{1}{\sqrt{6}}
\left( \vert\uparrow\downarrow\downarrow\rangle - 2\vert\downarrow\uparrow\downarrow\rangle + \vert\downarrow\downarrow\uparrow\rangle\right),
\qquad
\end{eqnarray}
i.e., the quadruplet $\vert q\rangle$ and the two doublets $\vert d1\rangle$, $\vert d2\rangle$.
The total spin of $\vert d1\rangle$ and  $\vert d2\rangle$ is 1/2 and the
`local'
$s_j^z$-values  for the sites $j=1,2,3$ are as follows:
$0,\,\pm 1/2,\,0$ for $\vert d1\rangle$ and 
$\pm1/3,\,\mp 1/6,\,\pm 1/3$ for $\vert d2\rangle$.
The states given in Eq.~(\ref{b03}) are the eigenstates of
$({\bf{s}}_{1}+{\bf{s}}_{2}+{\bf{s}}_{3})^2$, of $s_1^z+s_2^z+s_3^z$, 
and of the Hamiltonian $H={\bf{s}}_{1}\cdot{\bf{s}}_{2}+{\bf{s}}_{2}\cdot{\bf{s}}_{3}$
(three-site Heisenberg model with open boundary conditions) with the energies
$1/2$ ($\vert q\rangle$),
$0$ ($\vert d1\rangle$),
and
$-1$ ($\vert d2\rangle$).

Next we consider  ${\cal{N}}=4$ cells along a chain with open boundary
conditions and $n=4$ electrons. The unperturbed SU(2) symmetric ground states are 
\begin{eqnarray}
\label{b05}
\vert Q,2\rangle=\vert\uparrow\uparrow\uparrow\uparrow\rangle, 
\ldots ,
\vert Q,-2\rangle=\vert\downarrow\downarrow\downarrow\downarrow\rangle,
\nonumber\\
\vert t1,1\rangle
=
\frac{1}{2\sqrt{2-\sqrt{2}}}
\left[-\vert\uparrow\uparrow\uparrow\downarrow\rangle + (1-\sqrt{2})\vert\uparrow\uparrow\downarrow\uparrow\rangle
- (1-\sqrt{2})\vert\uparrow\downarrow\uparrow\uparrow\rangle + \vert\downarrow\uparrow\uparrow\uparrow\rangle\right],
\ldots ,
\nonumber\\
\vert t2,1\rangle
=
\frac{1}{2}\left(\vert\uparrow\uparrow\uparrow\downarrow\rangle - \vert\uparrow\uparrow\downarrow\uparrow\rangle\right)
+\frac{1}{2}\left(-\vert\uparrow\downarrow\uparrow\uparrow\rangle + \vert\downarrow\uparrow\uparrow\uparrow\rangle\right),
\ldots ,
\nonumber\\
\vert t3,1\rangle
=
\frac{1}{2\sqrt{2+\sqrt{2}}}
\left[-\vert\uparrow\uparrow\uparrow\downarrow\rangle + (1+\sqrt{2})\vert\uparrow\uparrow\downarrow\uparrow\rangle
-(1+\sqrt{2})\vert\uparrow\downarrow\uparrow\uparrow\rangle + \vert\downarrow\uparrow\uparrow\uparrow\rangle\right],
\ldots ,
\nonumber\\
\vert s1\rangle
=
\frac{1}{\sqrt{6}}
\left[
\frac{1}{2}\left(-1-\sqrt{3}\right)\vert\uparrow\uparrow\downarrow\downarrow\rangle 
- \frac{1}{2}\left(1-\sqrt{3}\right)\vert\uparrow\downarrow\uparrow\downarrow\rangle 
+ \vert\uparrow\downarrow\downarrow\uparrow\rangle
\right.
\nonumber\\
\left.
+ \vert\downarrow\uparrow\uparrow\downarrow\rangle
-\frac{1}{2}\left(1-\sqrt{3}\right)\vert\downarrow\uparrow\downarrow\uparrow\rangle 
+ \frac{1}{2}\left(-1-\sqrt{3}\right)\vert\downarrow\downarrow\uparrow\uparrow\rangle
\right],
\nonumber\\
\vert s2\rangle
=
\frac{1}{\sqrt{6}}
\left[
\frac{1}{2}\left(-1+\sqrt{3}\right)\vert\uparrow\uparrow\downarrow\downarrow\rangle 
- \frac{1}{2}\left(1+\sqrt{3}\right)\vert\uparrow\downarrow\uparrow\downarrow\rangle 
+ \vert\uparrow\downarrow\downarrow\uparrow\rangle
\right.
\nonumber\\
\left.
+ \vert\downarrow\uparrow\uparrow\downarrow\rangle
-\frac{1}{2}\left(1+\sqrt{3}\right)\vert\downarrow\uparrow\downarrow\uparrow\rangle 
+ \frac{1}{2}\left(-1+\sqrt{3}\right)\vert\downarrow\downarrow\uparrow\uparrow\rangle 
\right],
\end{eqnarray}
i.e., one quintuplet $\vert Q\rangle$, 
the three triplets $\vert t1\rangle$, $\vert t2\rangle$, $\vert t3\rangle$, 
and the two singlets $\vert s1\rangle$, $\vert s2\rangle$.
These  states are eigenstates of the Heisenberg Hamiltonian 
$H=\sum_{i=1}^{3}{\bf{s}}_{i}\cdot{\bf{s}}_{i+1}$ 
with the energies 
$3/4$ ($\vert Q\rangle$),
$(-1+2\sqrt{3})/4$ ($\vert t1\rangle$),
$-1/4$ ($\vert t2\rangle$),
$(-1-2\sqrt{3})/4$ ($\vert t3\rangle$),
$(-3+2\sqrt{3})/4$ ($\vert s1\rangle$),
and
$(-3-2\sqrt{3})/4$ ($\vert s2\rangle$).

In the case of ${\cal{N}}=5$ cells and $n=5$ electrons relevant for the bilayer
problem we have
\begin{eqnarray}
\label{b06}
\left\vert S,\frac{5}{2}\right\rangle=\vert\uparrow\uparrow\uparrow\uparrow\uparrow\rangle, 
\ldots ,
\left\vert S,-\frac{5}{2}\right\rangle=\vert\downarrow\downarrow\downarrow\downarrow\downarrow\rangle,
\nonumber\\
\left\vert q1,\frac{3}{2}\right\rangle
=
\frac{1}{2}
\left(
\vert\uparrow\uparrow\uparrow\uparrow\downarrow\rangle
-\vert\uparrow\uparrow\uparrow\downarrow\uparrow\rangle
+\vert\uparrow\downarrow\uparrow\uparrow\uparrow\rangle
-\vert\downarrow\uparrow\uparrow\uparrow\uparrow\rangle
\right),
\ldots ,
\nonumber\\
\left\vert q2,\frac{3}{2}\right\rangle
=
\frac{1}{2}
\left(
\vert\uparrow\uparrow\uparrow\uparrow\downarrow\rangle
-\vert\uparrow\uparrow\uparrow\downarrow\uparrow\rangle
-\vert\uparrow\downarrow\uparrow\uparrow\uparrow\rangle
+\vert\downarrow\uparrow\uparrow\uparrow\uparrow\rangle
\right),
\ldots ,
\nonumber\\
\left\vert q3,\frac{3}{2}\right\rangle
=
\frac{1}{2}
\left(
\vert\uparrow\uparrow\uparrow\uparrow\downarrow\rangle
+\vert\uparrow\uparrow\uparrow\downarrow\uparrow\rangle
-\vert\uparrow\downarrow\uparrow\uparrow\uparrow\rangle
-\vert\downarrow\uparrow\uparrow\uparrow\uparrow\rangle
\right),
\ldots ,
\nonumber\\
\left\vert q4,\frac{3}{2}\right\rangle
=
\frac{1}{2\sqrt{5}}
\left(
\vert\uparrow\uparrow\uparrow\uparrow\downarrow\rangle
+\vert\uparrow\uparrow\uparrow\downarrow\uparrow\rangle
-4\vert\uparrow\uparrow\downarrow\uparrow\uparrow\rangle
+\vert\uparrow\downarrow\uparrow\uparrow\uparrow\rangle
+\vert\downarrow\uparrow\uparrow\uparrow\uparrow\rangle
\right),
\ldots ,
\nonumber\\
\ldots ,
\end{eqnarray}
i.e.,
one sextuplet $\vert S\rangle$
and
the four quadruplets $\vert q1\rangle$, $\vert q2\rangle$, $\vert q3\rangle$, $\vert
q4\rangle$.
Note that the five doublets are not given here, since they are not used for perturbation theory, 
cf. the discussion in Sec.~\ref{sec6}. 
The geometry of the cluster is that of a Heisenberg star\cite{star} with
central spin ${\bf s}_3$, 
i.e.,
the choice given in Eq.~(\ref{b06}) corresponds
to the eigenstates of the Heisenberg Hamiltonian 
$H={\bf{s}}_{1}\cdot{\bf{s}}_{3}+{\bf{s}}_{2}\cdot{\bf{s}}_{3} +{\bf{s}}_{3}\cdot{\bf{s}}_{4}+{\bf{s}}_{3}\cdot{\bf{s}}_{5}$
with the energies 
$1$ ($\vert S\rangle$),
$1/2$ ($\vert q1\rangle$, $\vert q2\rangle$, and $\vert q3\rangle$),
and
$-3/2$ ($\vert q4\rangle$).

In the present study 
we use 
Eqs.~(\ref{b02}) and (\ref{b03}) for the diamond chain,
Eqs.~(\ref{b02}), (\ref{b03}), and (\ref{b05}) for the ladder,
and Eq.~(\ref{b06}) for the bilayer.
Since for the ${\cal{N}}=5$ bilayer we compare the energies $E_S$ and $E_{q1}$, $E_{q2}$, $E_{q3}$, $E_{q4}$ only,
the formulas given in Eq.~(\ref{b06}) are sufficient for this purpose.

\section*{Appendix C: Perturbation-theory results for the diamond chain}
\renewcommand{\theequation}{C\arabic{equation}}
\setcounter{equation}{0}

\subsection*{$n=2$ electrons on the diamond chain of ${\cal{N}}=2$ cells}

\begin{figure}%[h]
\begin{center}
\includegraphics[clip=on,width=80mm,angle=0]{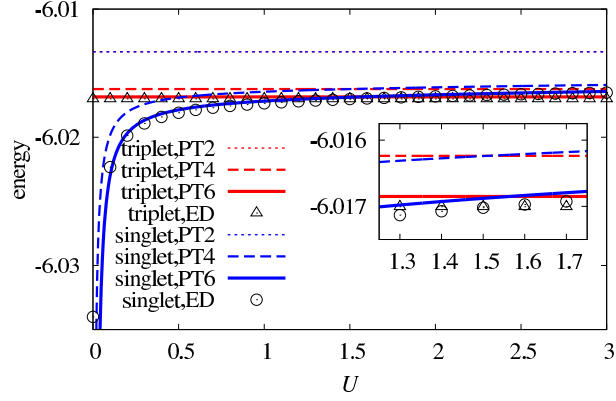}
\caption{(Color online)
Energies of low-lying states (triplet -- red, singlet -- blue) as a function of the on-site repulsion $U$
for $n=2$ electrons on ${\cal{N}}=2$ cells of the diamond chain with open boundary conditions 
with $t_2=3$, $t_{1}=0.9$, $t_{3}=1.1$.
The results up to the second, fourth, sixth orders are denoted by short-dashed, long-dashed, solid lines, respectively.
The results of exact diagonalization are shown by symbols. 
Note that the energies of the triplet and the singlet coincide within the second order, see Eq.~(\ref{c01}).}
\label{f08}
\end{center}
\end{figure}

We consider the case of open boundary conditions, i.e., $N=5$.
Corrections to the ground-state energy $E^{(0)}=-2t_{2}$ up to the sixth order are as follows:
\begin{eqnarray}
\label{c01}
E^{(2)}      =  -\frac{(t_{3}-t_{1})^{2}}{t_{2}};
\end{eqnarray}
\begin{eqnarray}
\label{c02}
E_{t}^{(4)}  = -\frac{(t_{3}+t_{1})^{2}(t_{3}-t_{1})^{2}}{2t_2^3}+\frac{(t_{3}-t_{1})^{4}}{t_2^3},
\nonumber\\
E_{s}^{(4)}(U)  = -\frac{\left(t_3+t_1\right)^2\left(t_3-t_1\right)^2}{4t_2^3}
+\frac{\left(t_3-t_1\right)^4}{t_2^3}
-\frac{\left(8t_2+U\right)\left(t_3-t_1\right)^4}{4t_2^3U}
-\frac{\left(t_3+t_1\right)^2 \left(t_3-t_1\right)^2}{2\left(2t_2+U\right)t_2^2}
-\frac{2\left(t_3-t_1\right)^4}{\left(2t_2+U\right)t_2^2};
\end{eqnarray}
\begin{eqnarray}
\label{c03}
E_{t}^{(6)}  =  -\frac{(t_{3}-t_{1})^{2}\left(t_{3}^{4}-14t_{3}^{3}t_{1}+34t_{3}^{2}t_{1}^{2}-14t_{3}t_{1}^{3}+t_{1}^{4}\right)}{2t_{2}^{5}},
\nonumber\\
E_{s}^{(6)}(U)  =  -\frac{(t_{3}-t_{1})^{2}\left[192t_2^{4}(t_{3}-t_{1})^{4}+48t_2^{3}U(t_{3}-t_{1})^{2}\left(5t_{3}^{2}-2t_{3}t_{1}+5t_{1}^{2}\right)\right]}{12t_{2}^{5}U^{2}(2t_{2}+U)^{2}}
\nonumber\\
-\frac{(t_{3}-t_{1})^{2}\left[4t_2^{2}U^{2}\left(22t_{3}^{4}+21t_{3}^{3}t_{1}-38t_{3}^{2}t_{1}^{2}+21t_{3}t_{1}^{3}+22t_{1}^{4}\right)\right]}{12t_{2}^{5}U^{2}(2t_{2}+U)^{2}}
\nonumber\\
-\frac{(t_{3}-t_{1})^{2}\left[4t_2U^{3}\left(2t_{3}^{4}+27t_{3}^{3}t_{1}-34t_{3}^{2}t_{1}^{2}+27t_{3}t_{1}^{3}+2t_{1}^{4}\right)\right]}{12t_{2}^{5}U^{2}(2t_{2}+U)^{2}}
\nonumber\\
-\frac{(t_{3}-t_{1})^{2}\left[U^{4}\left(t_{3}^{4}+12t_{3}^{3}t_{1}-14t_{3}^{2}t_{1}^{2}+12t_{3}t_{1}^{3}+t_{1}^{4}\right)\right]}{12t_{2}^{5}U^{2}(2t_{2}+U)^{2}}.
\end{eqnarray}
The results up to the fourth order were reported in Ref.~\onlinecite{oleg-johannes}.
In Fig.~\ref{f08} we show dependences of the triplet and singlet energies on $U$ 
obtained within different orders of the perturbation theory according to Eqs.~(\ref{c01}), (\ref{c02}), (\ref{c03}) along with exact-diagonalization data 
for a typical set of hopping integrals $t_2=3$, $t_1=0.9$, $t_3=1.1$ [$t=1$,
$\vert\delta\vert=0.1$, $\delta=(t_1-t_3)/(t_1+t_3)$].
Obviously,
in the limit $U\to 0$ the perturbation theory fails, since it yields a singlet energy tending to $-\infty$ whereas the exact-diagonalization data is finite.
The reason for that is clear:
Within the exploited scheme the specific states with two electrons having different spins in one cell are treated as excited states,
however, in the small-$U$ limit their energy approaches the ground-state energy;
being treated as excited states they lead to large denominators in the terms of the perturbation-theory
series,
see Eqs.~(\ref{a01}), (\ref{a02}).

\subsection*{$n=3$ electrons on the diamond chain of ${\cal{N}}=3$ cells}

\begin{figure}%[h]
\begin{center}
\includegraphics[clip=on,width=80mm,angle=0]{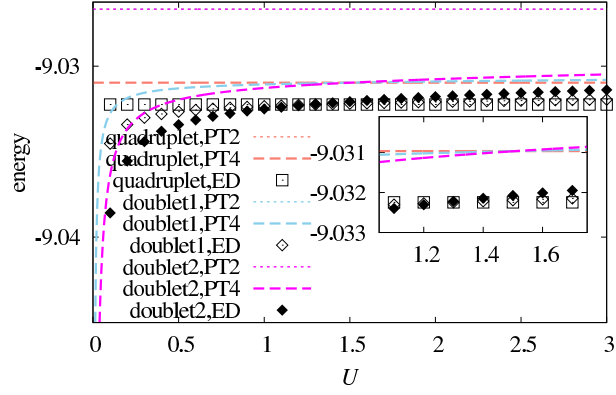}
\caption{(Color online)
Ground-state energy as a function of the on-site repulsion $U$  
for $n=3$ electrons on ${\cal{N}}=3$ cells of the diamond chain (open boundary conditions) with $t_2=3$, $t_{1}=0.9$, $t_{3}=1.1$.
Quadruplet energy (salmon) versus doublets energy (skyblue and magenta).
The results up to the second and fourth orders are denoted by short-dashed and long-dashed lines, respectively.
The results of exact diagonalization are shown by symbols. 
Note that the energies of the doublet and quadruplet states coincide within the second order, see Eq.~(\ref{c04}).}
\label{f09}
\end{center}
\end{figure}

We consider the case of open boundary conditions, i.e., $N=8$.
Corrections to the ground-state energy $E^{(0)}=-3t_{2}$ up to the fourth order are as follows:
\begin{eqnarray}
\label{c04}
E^{(2)}=-\frac{2(t_{1}-t_{3})^{2}}{t_{2}};
\end{eqnarray}
\begin{eqnarray}
\label{c05}
E_{q}^{(4)}    
=  
\frac{(t_{1}-t_{3})^{2}\left(7t_{1}^{2}-26t_{1}t_{3}+7t_{3}^{2}\right)}{4t_{2}^{3}},
\nonumber\\
E_{d1}^{(4)}(U)  
=  
\frac{(t_{1}-t_{3})^{2}\left[-24t_{2}^{2}(t_{1}-t_{3})^{2}+t_{2}U\left(t_{1}^{2}-50t_{1}t_{3}+t_{3}^{2}\right)+2U^{2}\left(7t_{1}^{2}-23t_{1}t_{3}+7t_{3}^{2}\right)\right]}{8t_{2}^{3}U(2t_{2}+U)},
\nonumber\\
E_{d2}^{(4)}(U)  
=  
\frac{(t_{1}-t_{3})^{2}\left[-40t_{2}^{2}(t_{1}-t_{3})^{2}-t_{2}U\left(17t_{1}^{2}+14t_{1}t_{3}+17t_{3}^{2}\right)+14U^{2}\left(t_{1}^{2}-3t_{1}t_{3}+t_{3}^{2}\right)\right]}{8t_{2}^{3}U(2t_{2}+U)}.
\end{eqnarray}
Splitting of various SU(2) multiplets begins in the fourth order of perturbation theory.
In Fig.~\ref{f09} we show dependences of the quadruplet and doublets energies on $U$ 
obtained within different orders of the perturbation theory according to Eqs.~(\ref{c04}), (\ref{c05})
along with exact-diagonalization data 
for the same set of hopping integrals as in Fig.~\ref{f08},
i.e., $t_2=3$, $t_1=0.9$, $t_3=1.1$.
At a first glance one may be worry about the agreement between perturbation theory and exact diagonalization.
However, 
comparing the fourth-order results and the exact-diagonalization data for ${\cal{N}}=2$ cells shown in Fig.~\ref{f08}
one can see a similar difference which is obviously improved by the the sixth-order calculations.

\section*{Appendix D: Perturbation-theory results for the ladder}
\renewcommand{\theequation}{D\arabic{equation}}
\setcounter{equation}{0}

\subsection*{$n=2$ electrons on the ladder of ${\cal{N}}=2$ cells}

For the two-cell ($N=4$) ladder (open boundary conditions)   
we have the following corrections to the unperturbed ground-state energy $E^{(0)}=-2t_2$:
\begin{eqnarray}
\label{d01}
E_t^{(2)}=-\frac{\left(t_{11}-t_{22}\right)^2+\left(t_{12}-t_{21}\right)^2}{4t_2},
\nonumber\\
E_s^{(2)}(U)=-\frac{\left(t_{11}-t_{12}-t_{21}+t_{22}\right)^2}{4t_2}-\frac{\left(t_{11}-t_{22}\right)^2+\left(t_{12}-t_{21}\right)^2}{2\left(2t_2+U\right)}
-\frac{2\left(t_{11}-t_{12} - t_{21}+t_{22}\right)^2}{U};
\end{eqnarray}
\begin{eqnarray}
\label{d02}
E_t^{(4)}
=
\frac{1}{64t_{2}^3}
\left[t_{11}^4-4t_{11}^3t_{22}+2t_{11}^2\left(3t_{22}^2-t_{21}^2-6t_{21}t_{12}-t_{12}^2\right) -4t_{11}t_{22}\left(t_{22}^2+3t_{21}^2-14t_{21}t_{12}+3t_{12}^2\right)
\right.
\nonumber\\
\left.
+t_{22}^4-2t_{22}^2\left(t_{21}^2+6t_{21}t_{12}+t_{12}^2\right)+\left(t_{21}-t_{12}\right)^4\right],
\nonumber\\
E_s^{(4)}(U)
=
\frac{1}{64t_2^3U^3\left(2t_2+U\right)^3}
\left\{
4096 t_2^6\left(t_{11}-t_{12}-t_{21}+t_{22}\right)^4 
+7680 t_2^5U \left(t_{11}-t_{12}-t_{21}+t_{22}\right)^4 
\right.
\nonumber\\
\left.
+t_{11}\left[67t_{22}^2-132 t_{22}\left(t_{12}+t_{21}\right)+66\left(t_{12}+t_{21}\right)^2\right]
+21t_{22}^3
-66t_{22}^2\left(t_{12}+t_{21}\right)
+66t_{22}\left(t_{12}+t_{21}\right)^2
\right.
\nonumber\\
\left.
+256t_2^4U^2\left(t_{11}-t_{12}-t_{21}+t_{22}\right)
\left[21t_{11}^3+t_{11}^2\left(67t_{22}-66\left(t_{12}+t_{21}\right)\right)
-\left(t_{12}+t_{21}\right)\left(21t_{21}^2+46t_{12}t_{21}+21t_{12}^2\right)\right]
\right.
\nonumber\\
\left.
+t_{11}^2\left[406t_{22}^2-784t_{22}\left(t_{12}+t_{21}\right)+386\left(t_{12}+t_{21}\right)^2\right]
\right.
\nonumber\\
\left.
+32t_2^3U^3\left[
55t_{22}^4-248t_{22}^3\left(t_{12}+t_{21}\right)+386t_{22}^2\left(t_{12}+t_{21}\right)^2
-8t_{22}\left(t_{12}+t_{21}\right)\left(31t_{21}^2+57t_{12}t_{21}+31t_{12}^2\right)
\right]
\right.
\nonumber\\
\left.
+32t_2^3U^3\left(55t^4_{21}+256t_{21}^3t_{12}+406t_{12}^2t_{21}^2+256t_{21}t_{12}^3+55t_{12}^4\right)
\right.
\nonumber\\
\left.
+32t_2^3U^34t_{11}\left[-2\left(t_{12}+t_{21}\right)\left(31t_{21}^2+67t_{12}t_{21}+31t_{12}^2\right)\right]
\right.
\nonumber\\
\left.
+32t_2^3U^3
\left[
55t^4_{11}+8t_{11}^3\left(32t_{22 }-31\left(t_{12}+t_{21}\right)\right)
\right.
\right.
\nonumber\\
\left.
\left.
+4t_{11}\left[64t_{22}^3-196t_{22}^2\left(t_{12}+t_{21}\right)+t_{22}\left(193t_{21}^2+392t_{12}t_{21}+193t_{12}^2\right)\right]
\right.
\right.
\nonumber\\
\left.
\left.
+t_{11}^2\left[156t_{22}^2-311t_{22}\left(t_{12}+t_{21}\right)+4\left(38t_{21}^2+77t_{12}t_{21}+38t_{12}^2\right)\right]
\right.
\right.
\nonumber\\
\left.
\left.
+t_{11}\left[100t^3_{22}-311t_{22}^2\left(t_{12}+t_{21}\right)+4t_{22}\left(77t^2_{21}+158t_{21}t_{12}+77t_{12}^2\right)
-\left(t_{12}+t_{21}\right)\left(97t_{21}^2+214t_{12}t_{21}+97t^2_{12}\right)
\right]
\right.
\right.
\nonumber\\
\left.
\left.
+20t_{22}^4
-97t^3_{22}\left(t_{12}+t_{21}\right)+4t^2_{22}\left(38t_{21}^2+77t_{12}t_{21}+38t_{12}^2\right)
-t_{22}\left(t_{12}+t_{21}\right)\left(97t^2_{21}+214t_{12}t_{21}+97t_{12}^2\right)
\right]
\right.
\nonumber\\
\left.
+16t_2^2U^4\left[20t_{11}^4+t_{11}^3\left(100t_{22}-97\left(t_{12}+t_{21}\right)\right)
+4\left(5t_{21}^4+25t^3_{21}t_{12}+39t_{21}^2t_{12}^2+25t_{21}t_{12}^3+5t_{12}^4\right)\right]
\right.
\nonumber\\
\left.
+2t_2U^5\left(t_{11}-t_{12}-t_{21}+t_{22}\right)
\left[23t_{22}^3-69t_{22}^2\left(t_{12}+t_{21}\right)+69t_{22}\left(t_{12}+t_{21}\right)^2-\left(t_{12}+t_{21}\right)\left(23t_{21}^2+30t_{12}t_{21}+23t_{12}^2\right)\right]
\right.
\nonumber\\
\left.
+2t_2U^5\left(t_{11}-t_{12}-t_{21}+t_{22}\right)
\left[23t_{11}^3+t_{11}^2\left(53t_{22}-69\left(t_{12}+t_{21}\right)\right)
+t_{11}\left( 53t_{22}^2-138t_{22}\left(t_{12}+t_{21}\right) +69\left(t_{12}+t_{21}\right)^2 \right)\right]
\right.
\nonumber\\
\left.
+U^6\left(t_{11}-t_{12}-t_{21}+t_{22}\right)^2
\left[
t_{11}^2+2t_{11}\left(t_{22}-3\left(t_{12}+t_{21}\right)\right)+t_{22}^2
-6t_{22}\left(t_{12}+t_{21}\right)+\left(t_{12}+t_{21}\right)^2
\right]
\right\}. \quad \quad
\end{eqnarray}
The formulas for the sixth-order corrections are too lengthy to be presented here,
although we use these formulas to produce the results reported in Figs.~\ref{f03}(a), \ref{f03}(b), and \ref{f03}(c).
The formulas in (\ref{d01}), (\ref{d02}) become simpler in two particular cases introduced in Sec.~\ref{sec5}.
For the symmetric deformation we have 
\begin{eqnarray}
\label{d03}
E_t^{(2)}=0,
\nonumber\\
E_s^{(2)}(U)=-\frac{\left(t_{11}-t_{12}\right)^2\left(8t_2+U\right)}{t_2U};
\end{eqnarray}
\begin{eqnarray}
\label{d04}
E_t^{(4)}=0,
\nonumber\\
E_s^{(4)}(U)=\frac{\left(t_{11}-t_{12}\right)^2
\left[
512t_2^3\left(t_{11}-t_{12}\right)^2+192t_2^2U\left(t_{11}-t_{12}\right)^2+32t_2U^2\left(t_{11}-t_{12}\right)^2+U^3\left(t_{11}^2-6t_{11}t_{12}+t_{12}^2\right)
\right]}{4t_2^3U^3}.
\end{eqnarray}
For the semi-symmetric deformation we have
\begin{eqnarray}
\label{d05}
E_t^{(2)}=-\frac{\left(t_{11}-t_{21}\right)^2}{2t_2},
\nonumber\\
E_s^{(2)}(U)=-\frac{\left(t_{11}-t_{21}\right)^2}{2t_2+U};
\end{eqnarray}
\begin{eqnarray}
\label{d06}
E_t^{(4)}=-\frac{t_{11}t_{21}\left(t_{11}-t_{21}\right)^2}{2t_2^3},
\nonumber\\
E_s^{(4)}(U)=-\frac{\left(t_{11}-t_{21}\right)^2\left[8t_2t_{11}t_{21}+U\left(t_{11}+t_{21}\right)^2\right]}{2t_2\left(2t_2+U\right)^3}.
\end{eqnarray}
Furthermore, the sixth-order corrections are as follows:
\begin{eqnarray}
\label{d07}
E_t^{(6)}=0,
\nonumber\\
E_s^{(6)}(U)
=
-\frac{1}{8 t_{2}^5 U^5 \left(16 t_{2}^2+U^2\right)}
\left[
(t_{11}-t_{12})^2 
\left(
524288 t_{2}^7 (t_{11}-t_{12})^4+327680 t_{2}^6 U(t_{11}-t_{12})^4 +131072 t_{2}^5 U^2 (t_{11}-t_{12})^4
\right.
\right.
\nonumber\\
\left.
\left.
+1024 t_{2}^4 U^3(t_{11}-t_{12})^2 \left(35 t_{11}^2-76 t_{11} t_{12}+35 t_{12}^2\right)
+256 t_{2}^3 U^4 (t_{11}-t_{12})^2 \left(28 t_{11}^2-65 t_{11}t_{12}+28 t_{12}^2\right)
\right.
\right.
\nonumber\\
\left.
\left.
+8 t_{2}^2 U^5 \left(121 t_{11}^4-556 t_{11}^3t_{12}+886 t_{11}^2 t_{12}^2-556 t_{11} t_{12}^3+121 t_{12}^4\right)
+16 t_{2}U^6 (t_{11}-t_{12})^2 \left(4 t_{11}^2-17 t_{11} t_{12}+4t_{12}^2\right)
\right.
\right.
\nonumber\\
\left.
\left.
+U^7 \left(t_{11}^4-14 t_{11}^3 t_{12}+34 t_{11}^2t_{12}^2-14 t_{11} t_{12}^3+t_{12}^4\right)
\right)
\right] \qquad   
\end{eqnarray}
(symmetric deformation)
and
\begin{eqnarray}
\label{d08}
E_t^{(6)}
=
\frac{\left(t_{11}-t_{21}\right)^2\left(t_{21}^4+4t_{21}^3t_{11}-26t_{21}^2t_{11}^2+4t_{21}t_{11}^3+t_{11}^4\right)}{32t_2^5},
\nonumber\\
E_s^{(6)}(U)
=
\frac{\left(t_{11}-t_{21}\right)^2 \left(t^4_{21}+4t_{21}^3t_{11}-26t_{21}^2t_{11}^2+4t_{21}t_{11}^3+t_{11}^4\right)}{\left(2t_2+U\right)^5}
\nonumber\\
+\frac{\left(t_{11}-t_{21}\right)^2\left[t_2U\left(t_{11}+t_{21}\right)^2\left(3t_{21}^2-14t_{21}t_{11}+3t_{11}^2\right)
-2t_{21}t_{11}U^2  \left(t_{11}+t_{21}\right)^2 \right]}{2t_2^2\left(2t_2+U\right)^5}
\end{eqnarray}
(semi-symmetric deformation).

\subsection*{$n=3$ electrons on the ladder of ${\cal{N}}=3$ cells}

\begin{figure}%[h]
\begin{center}
\includegraphics[clip=on,width=80mm,angle=0]{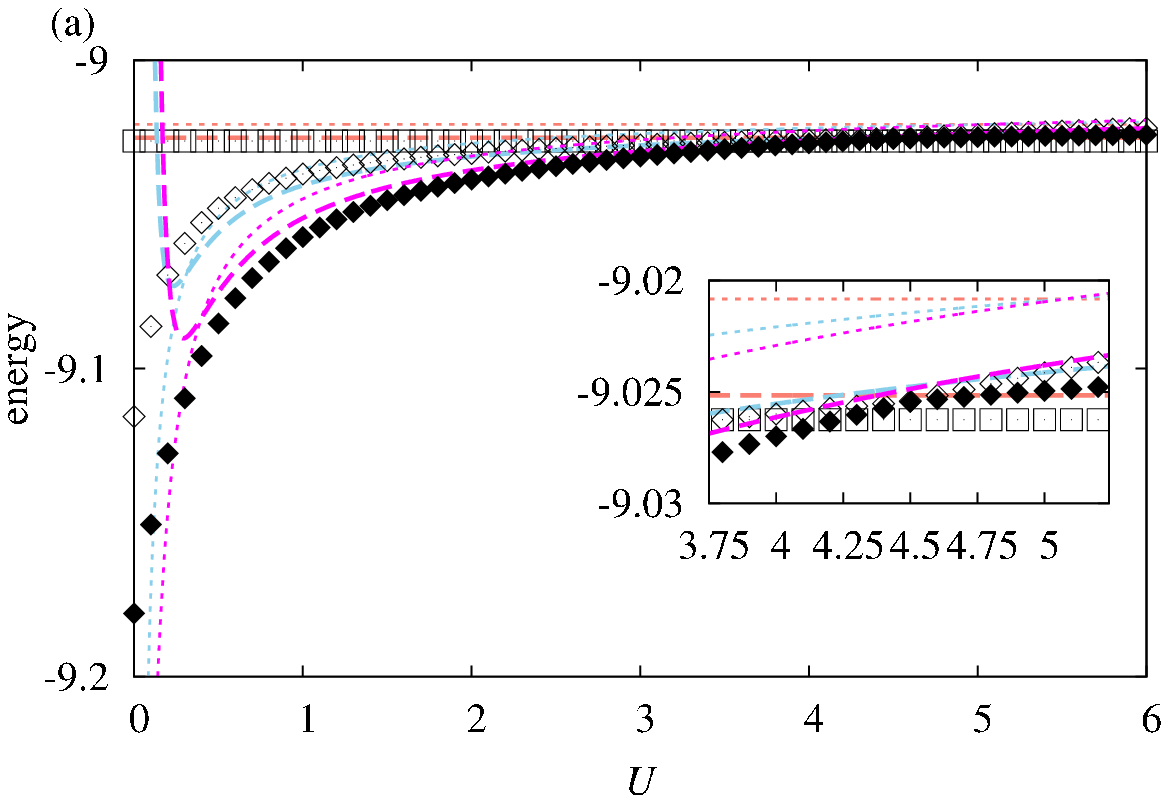}\\
\includegraphics[clip=on,width=80mm,angle=0]{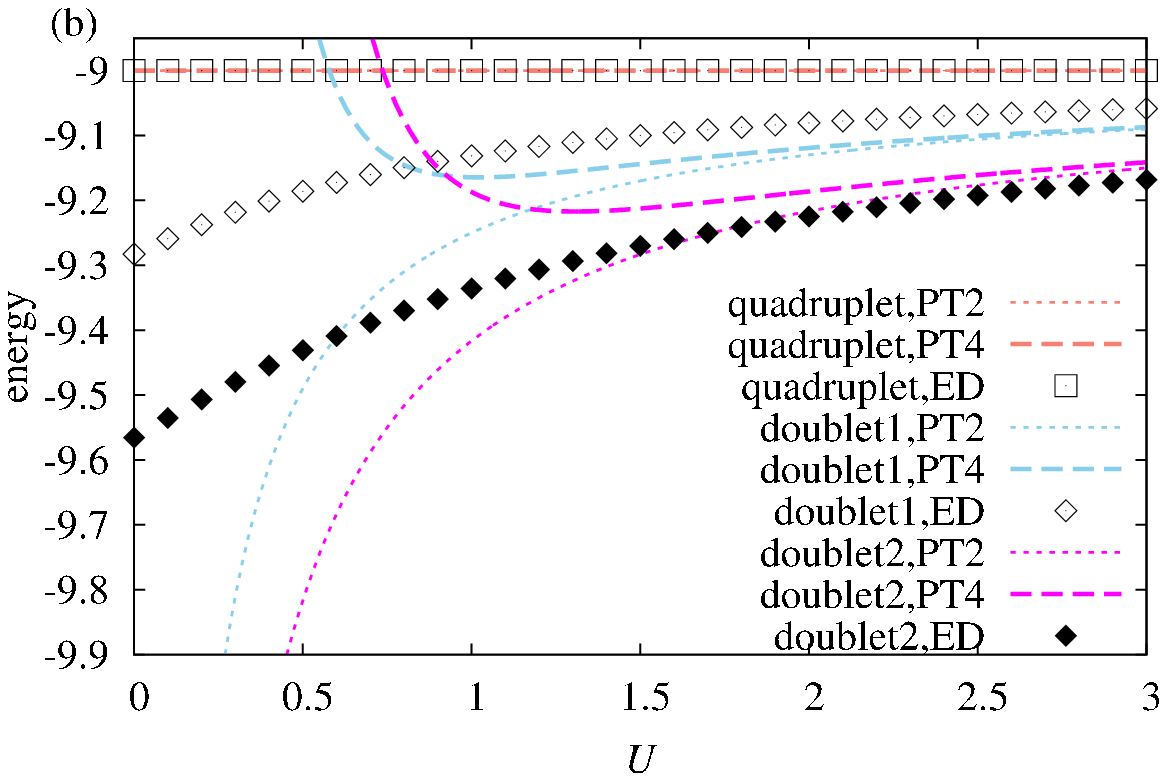}\\
\includegraphics[clip=on,width=80mm,angle=0]{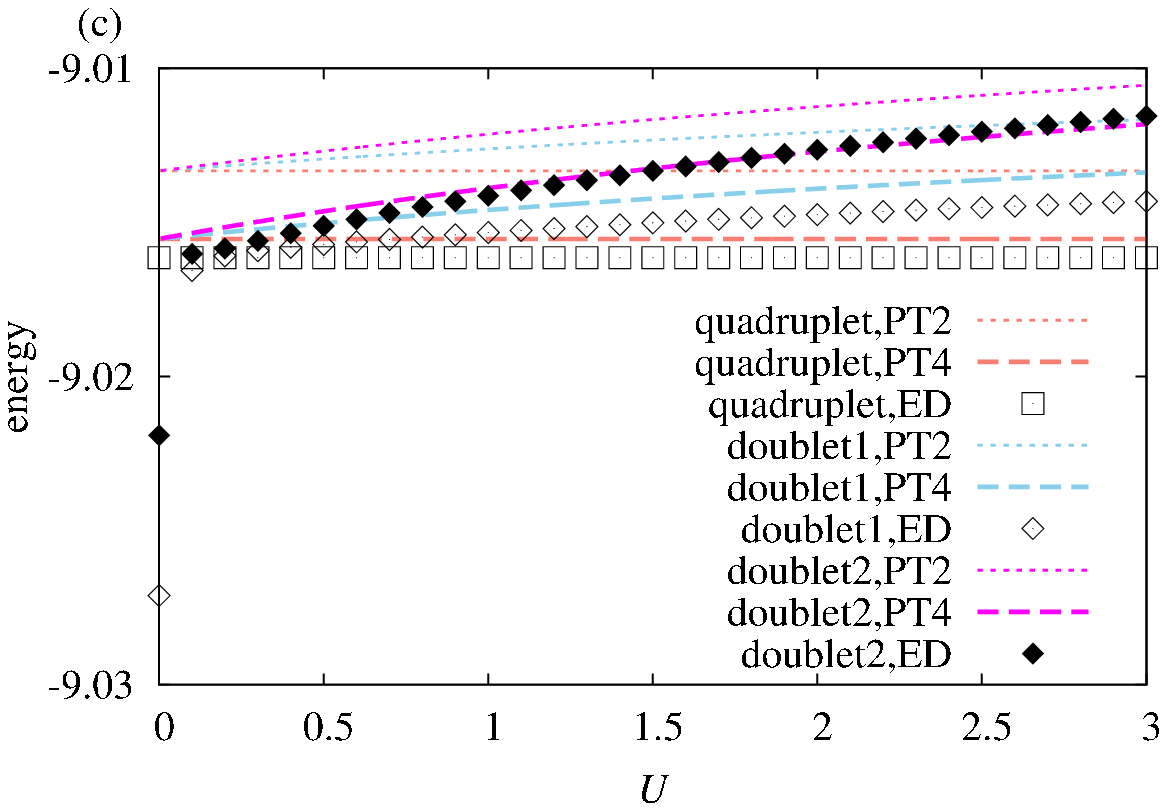} 
\caption{(Color online)
Ground-state energy (up to the fourth order of perturbation theory and exact-diagonalization data) 
as a function
of the on-site repulsion $U$
for $n=3$ electrons on the open ladder of ${\cal{N}}=3$ cells. 
(a)
$t_2=3$, $t_{11}=0.85$, $t_{12}=0.95$, $t_{21}=1$, $t_{22}=1.2$ (general deformation).
(b)
$t_2=3$, $t_{11}=t_{22}=1.1$, $t_{12}=t_{21}=0.9$ (symmetric deformation).
(c)
$t_2=3$, $t_{11}=t_{21}=1.1$, $t_{12}=t_{22}=0.9$ (semi-symmetric deformation).}
\label{f10}
\end{center}
\end{figure}

For the three-cell ($N=6$) ladder  (open boundary conditions)
we have the following corrections to the unperturbed ground-state energy $E^{(0)}=-3t_2$:
\begin{eqnarray}
\label{d09}
E_{q}^{(2)}  
= 
-\frac{(t_{11}-t_{22})^2+(t_{12}-t_{21})^2}{2 t_{2}},
\nonumber\\
E_{d1}^{(2)}(U)
= 
\frac{1}{4} 
\left(
\frac{-2 t_{11}^2+t_{11} (2 t_{22}+t_{12}+t_{21})-2 t_{22}^2+t_{22} (t_{12}+t_{21})-2\left(t_{12}^2-t_{12} t_{21}+t_{21}^2\right)}{t_{2}}
\right.
\nonumber\\
\left.
-\frac{(t_{11}-t_{22})^2+(t_{12}-t_{21})^2}{2 t_{2}+U}
-\frac{4(t_{11}+t_{22}-t_{12}-t_{21})^2}{U}
\right),
\nonumber\\
E_{d2}^{(2)}(U)
= 
-\frac{1}{4 t_{2} U (2 t_{2}+U)}
\left[
(8 t_{2}+U) \left(3 t_{2} (t_{11}+t_{22}-t_{12}-t_{21})^2
\right.
\right.
\nonumber\\
\left.
\left.
+U \left(2 t_{11}^2+t_{11} (2 t_{22}-3 (t_{12}+t_{21}))+2 t_{22}^2
-3 t_{22} (t_{12}+t_{21})+2 \left(t_{12}^2+t_{12} t_{21}+t_{21}^2\right)\right)\right)
 \right];
\end{eqnarray}
\begin{eqnarray}
\label{d10}
E_{q}^{(4)}  
= 
\frac{1}{16 t_{2}^3}
\left[
t_{11}^4-4 t_{11}^3 t_{22}+t_{11}^2 \left(6 t_{22}^2-3 t_{12}^2-10 t_{12} t_{21}-3 t_{21}^2\right)
\right.
\nonumber\\
\left.
+2 t_{11} t_{22}
   \left(-2 t_{22}^2+t_{12}^2+14 t_{12} t_{21}+t_{21}^2\right)+t_{22}^4-t_{22}^2 \left(3 t_{12}^2+10 t_{12} t_{21}+3
   t_{21}^2\right)-4 t_{12} t_{21} (t_{12}-t_{21})^2
\right],
\end{eqnarray}
and the formulas for $E_{d1}^{(4)}(U)$ and $E_{d2}^{(4)}(U)$ are too lengthy to be presented here.
Formulas given in Eqs.~(\ref{d09}), (\ref{d10}) are illustrated in
Fig.~\ref{f10},
where we show the dependence of energies the quadruplet and doublets  on $U$ for three typical sets of parameters.

\subsection*{$n=4$ electrons on the ladder of ${\cal{N}}=4$ cells}

\begin{figure}%[h]
\begin{center}
\includegraphics[clip=on,width=80mm,angle=0]{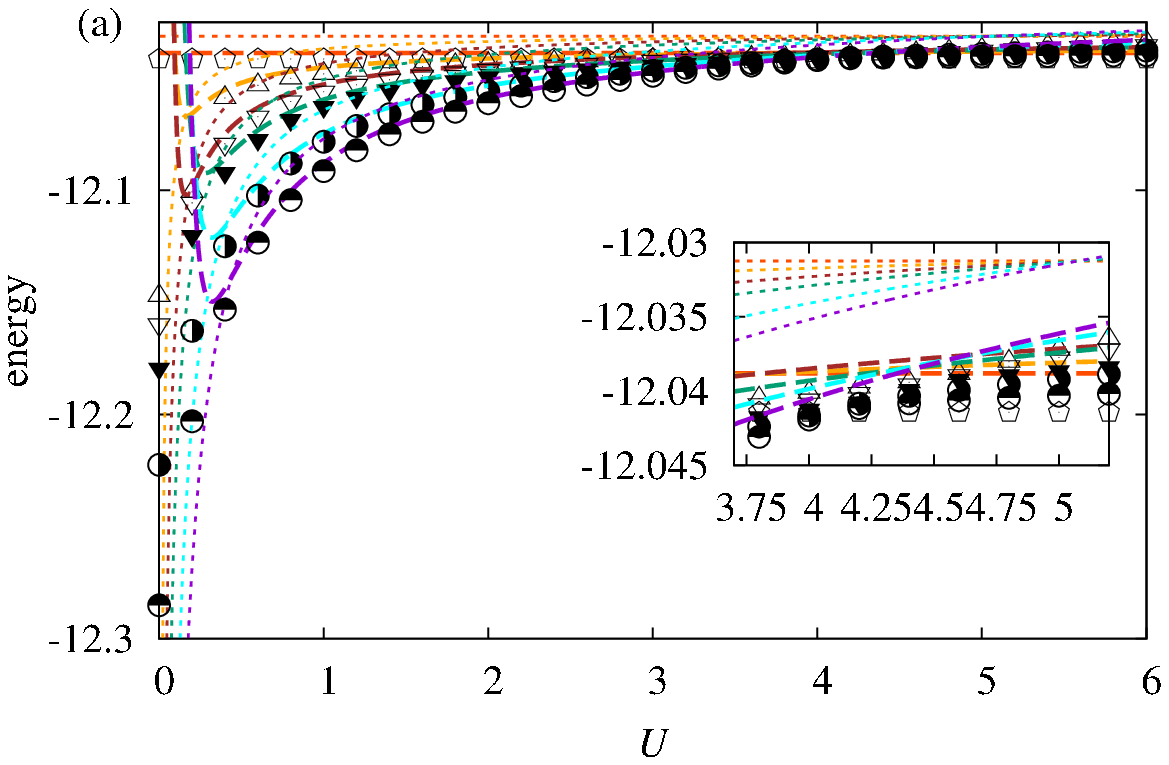}\\
\includegraphics[clip=on,width=80mm,angle=0]{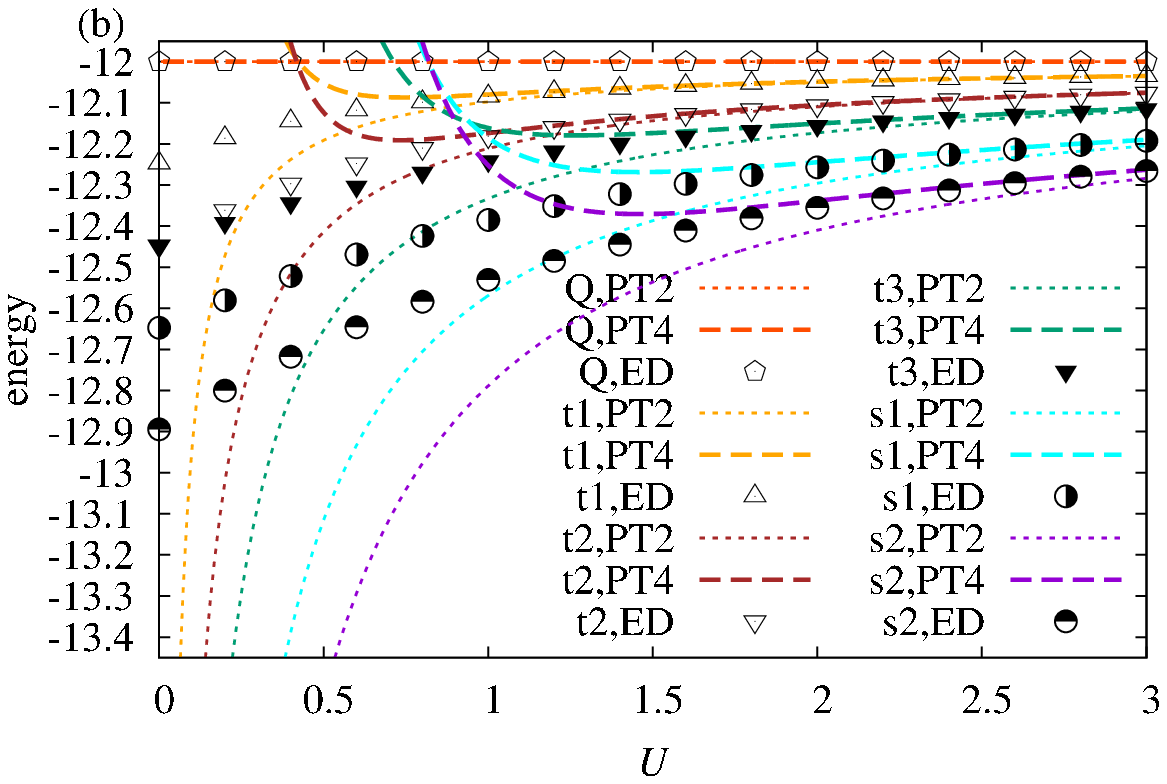}\\
\includegraphics[clip=on,width=80mm,angle=0]{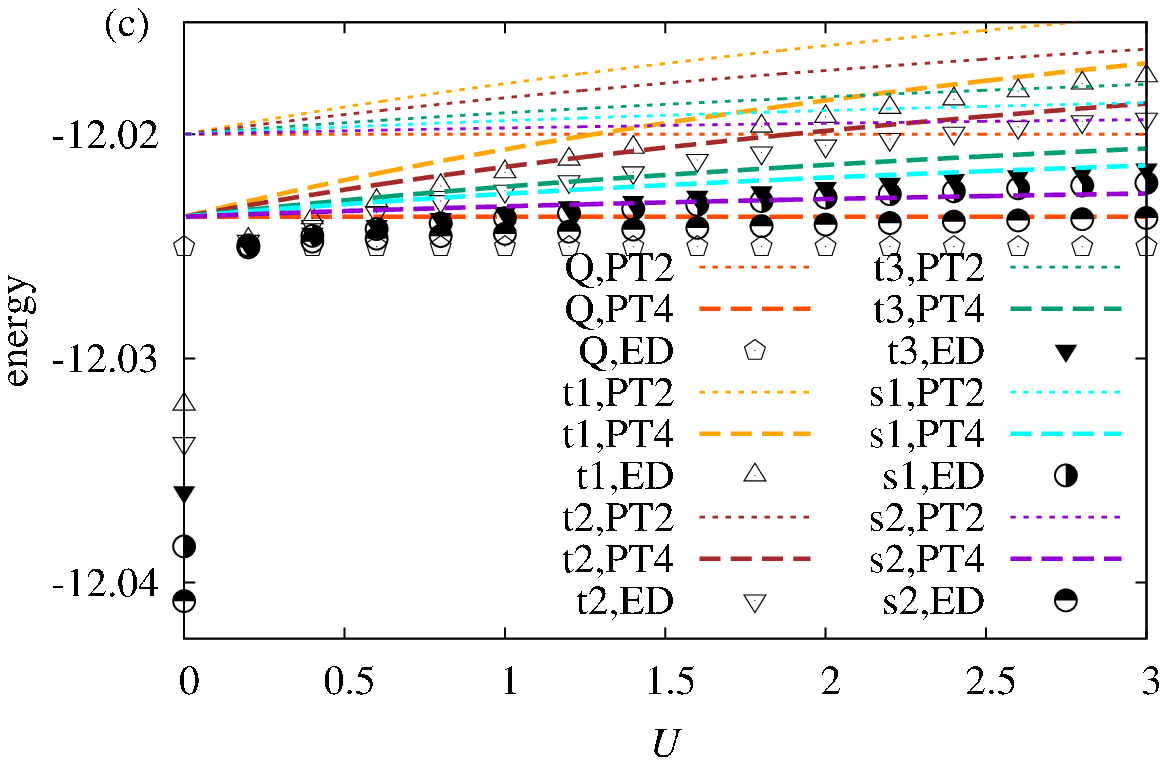} 
\caption{(Color online)
Ground-state energy (up to the fourth order of perturbation theory and exact-diagonalization data) 
as a function
of the on-site repulsion $U$
for $n=4$ electrons on the open ladder of ${\cal{N}}=4$ cells. 
(a)
$t_2=3$, $t_{11}=0.85$, $t_{12}=0.95$, $t_{21}=1$, $t_{22}=1.2$ (general deformation).
(b)
$t_2=3$, $t_{11}=t_{22}=1.1$, $t_{12}=t_{21}=0.9$ (symmetric deformation).
(c)
$t_2=3$, $t_{11}=t_{21}=1.1$, $t_{12}=t_{22}=0.9$ (semi-symmetric deformation).}
\label{f11}
\end{center}
\end{figure}

For the four-cell ($N=8$) ladder (open boundary conditions)   
we have the following corrections to the unperturbed ground-state energy $E^{(0)}=-4t_2$:
\begin{eqnarray}
\label{d11}
E_{Q}^{(2)}  = -\frac{3\left((t_{11}-t_{22})^{2}+(t_{12}-t_{21})^{2}\right)}{4t_{2}},
\nonumber\\
E_{t1}^{(2)}(U)
=  
\frac{1}{4\left(\sqrt{2}-2\right)t_{2}U(2t_{2}+U)}
\left[
2t_{2}U\left(\left(21-13\sqrt{2}\right)t_{11}^{2}+2t_{11}\left(\left(9-7\sqrt{2}\right)t_{22}+5\left(2\sqrt{2}-3\right)(t_{12}+t_{21})\right)
\right.
\right.
\nonumber\\
\left.
\left.
+\left(21-13\sqrt{2}\right)t_{22}^{2}+10\left(2\sqrt{2}-3\right)t_{22}(t_{12}+t_{21})+\left(21-13\sqrt{2}\right)t_{12}^{2}+2\left(9-7\sqrt{2}\right)t_{12}t_{21}+\left(21-13\sqrt{2}\right)t_{21}^{2}\right)
\right]
\nonumber\\
-\frac{1}{4\left(\sqrt{2}-2\right)t_{2}U(2t_{2}+U)}
\left[
U^{2}\left(3\left(\sqrt{2}-2\right)t_{11}^{2}+2t_{11}\left(\sqrt{2}t_{22}+\left(3-2\sqrt{2}\right)(t_{12}+t_{21})\right)+3\left(\sqrt{2}-2\right)t_{22}^{2}
\right.
\right.
\nonumber\\
\left.
\left.
+\left(6-4\sqrt{2}\right)t_{22}(t_{12}+t_{21})+3\left(\sqrt{2}-2\right)t_{12}^{2}+2\sqrt{2}t_{12}t_{21}+3\left(\sqrt{2}-2\right)t_{21}^{2}\right)
\right]
\nonumber\\
-\frac{16\left(2\sqrt{2}-3\right)t_{2}^{2}(t_{11}+t_{22}-t_{12}-t_{21})^{2}}{4\left(\sqrt{2}-2\right)t_{2}U(2t_{2}+U)},
\nonumber\\
E_{t2}^{(2)}(U)
=
\frac{1}{4}
\left(\frac{-3t_{11}^{2}+2t_{11}(t_{22}+t_{12}+t_{21})-3t_{22}^{2}+2t_{22}(t_{12}+t_{21})-3t_{12}^{2}+2t_{12}t_{21}-3t_{21}^{2}}{t_{2}}
\right.
\nonumber\\
\left.
-\frac{2\left((t_{11}-t_{22})^{2}+(t_{12}-t_{21})^{2}\right)}{2t_{2}+U}-\frac{8(t_{11}+t_{22}-t_{12}-t_{21})^{2}}{U}
\right),
\nonumber\\
E_{t3}^{(2)}(U)
=
-\frac{1}{4\left(2+\sqrt{2}\right)t_{2}U(2t_{2}+U)}
\left[2t_{2}U\left(\left(21+13\sqrt{2}\right)t_{11}^{2}+2t_{11}\left(\left(9+7\sqrt{2}\right)t_{22}-5\left(3+2\sqrt{2}\right)(t_{12}+t_{21})\right)
\right.
\right.
\nonumber\\
\left.
\left.
+\left(21+13\sqrt{2}\right)t_{22}^{2}-10\left(3+2\sqrt{2}\right)t_{22}(t_{12}+t_{21})+\left(21+13\sqrt{2}\right)t_{12}^{2}+2\left(9+7\sqrt{2}\right)t_{12}t_{21}+\left(21+13\sqrt{2}\right)t_{21}^{2}\right)
\right]
\nonumber\\
-\frac{1}{4\left(2+\sqrt{2}\right)t_{2}U(2t_{2}+U)}
\left[
U^{2}\left(3\left(2+\sqrt{2}\right)t_{11}^{2}+2t_{11}\left(\sqrt{2}t_{22}-\left(3+2\sqrt{2}\right)(t_{12}+t_{21})\right)+3\left(2+\sqrt{2}\right)t_{22}^{2}
\right.
\right.
\nonumber\\
\left.
\left.
-2\left(3+2\sqrt{2}\right)t_{22}(t_{12}+t_{21})+3\left(2+\sqrt{2}\right)t_{12}^{2}+2\sqrt{2}t_{12}t_{21}+3\left(2+\sqrt{2}\right)t_{21}^{2}\right)
\right]
\nonumber\\
- \frac{16\left(3+2\sqrt{2}\right)t_{2}^{2}(t_{11}+t_{22}-t_{12}-t_{21})^{2}}{4\left(2+\sqrt{2}\right)t_{2}U(2t_{2}+U)},
\nonumber\\
E_{s1}^{(2)}(U)
=
\frac{1}{4}
\left(
-\frac{3t_{11}^{2}+t_{11}\left(\left(\sqrt{3}-3\right)(t_{12}+t_{21})-2\sqrt{3}t_{22}\right)+3t_{22}^{2}+\left(\sqrt{3}-3\right)t_{22}(t_{12}+t_{21})+3t_{12}^{2}-2\sqrt{3}t_{12}t_{21}+3t_{21}^{2}}{t_{2}}
\right.
\nonumber\\
\left.
+\frac{\left(\sqrt{3}-3\right)\left((t_{11}-t_{22})^{2}+(t_{12}-t_{21})^{2}\right)}{2t_{2}+U}+\frac{4\left(\sqrt{3}-3\right)(t_{11}+t_{22}-t_{12}-t_{21})^{2}}{U}
\right),
\nonumber\\
E_{s2}^{(2)}(U)
= 
\frac{1}{4}\left(
\frac{-3t_{11}^{2}+t_{11}\left(\left(3+\sqrt{3}\right)(t_{12}+t_{21})-2\sqrt{3}t_{22}\right)-3t_{22}^{2}+\left(3+\sqrt{3}\right)t_{22}(t_{12}+t_{21})-3t_{12}^{2}-2\sqrt{3}t_{12}t_{21}-3t_{21}^{2}}{t_{2}}
\right.
\nonumber\\
\left.
-\frac{\left(3+\sqrt{3}\right)\left((t_{11}-t_{22})^{2}+(t_{12}-t_{21})^{2}\right)}{2t_{2}+U}-\frac{4\left(3+\sqrt{3}\right)(t_{11}+t_{22}-t_{12}-t_{21})^{2}}{U}
\right);
\end{eqnarray}
\begin{eqnarray}
\label{d12}
E_{Q}^{(4)}  
= 
\frac{1}{{64t_{2}^{3}}}
\left[
7t_{11}^{4}-28t_{11}^{3}t_{22}+t_{11}^{2}\left(42t_{22}^{2}-22t_{12}^{2}-68t_{12}t_{21}-22t_{21}^{2}\right)
+28t_{11}t_{22}\left(-t_{22}^{2}+t_{12}^{2}+6t_{12}t_{21}+t_{21}^{2}\right)
\right.
\nonumber\\
\left.
+7t_{22}^{4}-2t_{22}^{2}\left(11t_{12}^{2}+34t_{12}t_{21}+11t_{21}^{2}\right)-(t_{12}-t_{21})^{2}\left(t_{12}^{2}+30t_{12}t_{21}+t_{21}^{2}\right)
\right],
\end{eqnarray}
and the formulas for $E_{t1}^{(4)}(U)$, $E_{t2}^{(4)}(U)$, $E_{t3}^{(4)}(U)$, $E_{s1}^{(4)}(U)$, and $E_{s2}^{(4)}(U)$ 
are too lengthy to be presented here.
In Fig.~\ref{f11} we illustrate the dependence of the quintuplet, triplets, and singlets energies on $U$ for three typical sets of parameters.

\section*{Appendix E: Perturbation-theory results for the bilayer}
\renewcommand{\theequation}{E\arabic{equation}}
\setcounter{equation}{0}

\subsection*{$n=5$ electrons on the bilayer of ${\cal{N}}=5$ cells}

For the finite-size bilayer cluster (star geometry) 
 we have obtained the following corrections to the unperturbed ground-state energy $E^{(0)}=-5t_2$:
\begin{eqnarray}
\label{e01}
E_{S}^{(2)}  
= 
-\frac{(t_{11}-t_{22})^{2}+(t_{12}-t_{21})^{2}}{t_{2}},
\nonumber\\
E_{q1}^{(2)}(U) = E_{q2}^{(2)}(U) = E_{q3}^{(2)}(U)
\nonumber\\
=
\frac{1}{16}\left(\frac{-16t_{11}^{2}+t_{11}(14t_{22}+9(t_{12}+t_{21}))-16t_{22}^{2}+9t_{22}(t_{12}+t_{21})-2\left(8t_{12}^{2}-7t_{12}t_{21}+8t_{21}^{2}\right)}{t_{2}}
\right.
\nonumber\\
\left.
-\frac{9\left((t_{11}-t_{22})^{2}+(t_{12}-t_{21})^{2}\right)}{2t_{2}+U}-\frac{36(t_{11}+t_{22}-t_{12}-t_{21})^{2}}{U}\right),
\nonumber\\
E_{q4}^{(2)}(U)
= 
\frac{1}{16}\left(
\frac{-16t_{11}^{2}+t_{11}(22t_{22}+5(t_{12}+t_{21}))-16t_{22}^{2}+5t_{22}(t_{12}+t_{21})-2\left(8t_{12}^{2}-11t_{12}t_{21}+8t_{21}^{2}\right)}{t_{2}}
\right.
\nonumber\\
\left.
-\frac{5\left((t_{11}-t_{22})^{2}+(t_{12}-t_{21})^{2}\right)}{2t_{2}+U}-\frac{20(t_{11}+t_{22}-t_{12}-t_{21})^{2}}{U}\right);
\end{eqnarray}
\begin{eqnarray}
\label{e02}
E_{S}^{(4)}  
= 
\frac{1}{4t_{2}^{3}}
\left[
t_{11}^{4}-4t_{11}^{3}t_{22}+2t_{11}^{2}\left(3t_{22}^{2}-t_{12}^{2}-6t_{12}t_{21}-t_{21}^{2}\right)
-4t_{11}t_{22}\left(t_{22}^{2}+3t_{12}^{2}-14t_{12}t_{21}+3t_{21}^{2}\right)
\right.
\nonumber\\
\left.
+t_{22}^{4}-2t_{22}^{2}\left(t_{12}^{2}+6t_{12}t_{21}+t_{21}^{2}\right)+(t_{12}-t_{21})^{4}
\right],
\end{eqnarray}
and the formulas for 
$E_{q1}^{(4)}(U)$, $E_{q2}^{(4)}(U)$,  $E_{q3}^{(4)}(U)$, and $E_{q4}^{(4)}(U)$ 
are too lengthy to be presented here.

\section*{Appendix F: Mathematica tutorial. $n=3$ electrons on the diamond chain of $N=8$ sites}

After installing and calling the SNEG package in a Mathematica sheet
one basically needs the commands outlined below. At first all appearing annihilation operators have to be defined  with the command
\[
\text{snegfermionoperators}(\text{c1},\text{c2},\ldots,\text{c8},\text{la},\text{lb},\ldots,\text{dc});
\]
The occurring numbers correspond to the lattice sites. 
We also define new operators $l_{A}=(c_{A,1}-c_{A,2})/\sqrt{2}$
[see Eq.~(\ref{301})], 
as well as $d_{A}=(c_{A,1}+c_{A,2})/\sqrt{2}$,
to describe the ground states of the unperturbed Hamiltonian. 
This is implemented with the definition of rules, which
are explaining the connection to the new set of operators:
{
\begin{eqnarray*}
\text{rules} & = & \text{snegold2newrules}\left(\{\text{c1}(),\ldots,\text{c8}()\},\{\text{la}(),\ldots,\text{dc}()\},\left\{ \frac{\text{c1}()-\text{c2}()}{\sqrt{2}},\ldots,\frac{\text{c7}()+\text{c8}()}{\sqrt{2}}\right\} \right)
\end{eqnarray*}
}

The unperturbed Hamiltonian, which corresponds to Eq.~(\ref{401}), 
is called with the command
\begin{eqnarray*}
H_{0} & = & t_{2}\left(\text{hop}(\text{c1}(),\text{c2}())+\text{hop}(\text{c4}(),\text{c5}())+\text{hop}(\text{c7}(),\text{c8}())\right)\\
 & + & U\mathbf{(\text{hubbard}(\text{c1}())+\text{hubbard}(\text{c2}())+\text{hubbard}(\text{c3}())+\text{hubbard}(\text{c4}()))}\\
 & + & U(\text{hubbard}(\text{c5}())+\text{hubbard}(\text{c6}())+\text{hubbard}(\text{c7}())+\text{hubbard}(\text{c8}()))
\end{eqnarray*}

The next step is solving the unperturbed system $H_{0}$. 
If one constructs the basis-set with, e.g.,
\[
\text{basis}=\text{qbasis}(\{\text{la}(),\ldots,\text{dc}(),\text{c3}(),\text{c6}()\})
\]
the Hamilton matrix is built with the command
\[
\text{ham}=\text{makematricesbzop}(H\text{/.}\,\text{rules},\text{basis}[[4\text{;;}4]]),
\]
where $\text{basis}[[4\text{;;}4]]$ chooses the filling $Q$ of the considered states. 
Here the filling corresponds to $Q=-5=-{\rm{sites}}+{\rm{electrons}}$.
To get the Eigenenergies and the Eigenfunctions of $H_{0}$ one needs to call
\begin{eqnarray*}
\text{hamSz0} & = & \text{Select}[\text{ham},\text{First}[\text{\ensuremath{\#}1}]=\{-5\}\&][[1,2]];\\
\text{values} & = & \text{maskOp}(\text{Eigensystem},\text{hamSz0});
\end{eqnarray*}

As discussed in the previous sections 
it is advisable to combine the ground-state manifold of the unperturbed Hamiltonian to respect the SU(2) symmetry.
Since the Eigenfunctions are stored in \textit{values}, a new set of Eigenfunctions is provided by, e.g.,
\[
\text{For}[i=1,i\leq\text{Length}[\text{values}[[1]]],i\text{++},\text{maskOp}(\text{NormedVecs}(i)=\text{values}[[2,i]])];
\]
\begin{eqnarray*}
 & \ldots\\
\text{NormedVecsSU2}(138) & = & \text{NormedVecs}(136)+\text{NormedVecs}(138)-2\text{NormedVecs}(139);\\
\text{NormedVecsSU2}(139) & = & \text{NormedVecs}(135)-\text{NormedVecs}(137);\\
 & \ldots
\end{eqnarray*}

The next step is to set up the perturbation part with
\begin{eqnarray*}
V & = & t_{1}(\text{hop}(\text{c6}(),\text{c8}())+\text{hop}(\text{c4}(),\text{c6}())+\text{hop}(\text{c3}(),\text{c5}())+\text{hop}(\text{c1}(),\text{c3}()))\\
 & + & t_{3}(\text{hop}(\text{c2}(),\text{c3}())+\text{hop}(\text{c3}(),\text{c4}())+\text{hop}(\text{c6}(),\text{c7}())+\text{hop}(\text{c5}(),\text{c6}()));\\
\text{Vsz} & = & \text{makematricesbzop}(V\text{/.}\,\text{rules},\text{basis}[[4\text{;;}4]]);\\
\text{vsz1} & = & \text{Select}[\text{Vsz},\text{First}[\text{\ensuremath{\#}1}]=\{-5\}\&][[1,2]];
\end{eqnarray*}
The Elements $V_{i,j}$ then are given by
\begin{eqnarray*}
\text{For}[i=1,i\leq\text{Length}[\text{values}[[1]]],i\text{++},\\
\text{For}[j=1,j\leq\text{Length}[\text{values}[[1]]],j\text{++,} & \text{ElementsV}(i,j)=(\text{vsz1}.\text{NormedVecsSU2}(j)).\text{NormedVecsSU2}(i)]]
\end{eqnarray*}

Finally, the energy corrections can be computed easily from the formulas (\ref{a01}). 
The energy correction of the first order, e.g., is given by $E_{GS}^{(1)}=\text{ElementsV}(GS,GS)$.

\end{document}